\documentclass{amsart}
\usepackage{graphicx}

\usepackage{enumitem}
\setlist[enumerate]{leftmargin=0.75in}
\setlist[itemize]{leftmargin=0.75in}
\usepackage{caption}
\usepackage{subcaption}
\usepackage{amsmath, amsfonts}
\usepackage{amsthm}
\usepackage[margin=1in]{geometry}
\usepackage{tikz}
\usepackage{pgfplots}
\usepackage{amssymb}
\usepackage{pifont}
\newcommand{\cmark}{\ding{51}}
\newcommand{\xmark}{\ding{55}}
\usepackage{array}
\usepackage{enumerate}
\usepackage{rotating}
\usepackage{todonotes}
\usepackage{colortbl}

\usepackage{comment}

\usepackage{url}
\usepackage{cite}
\usepackage{float}

\usepackage{float}
\usepackage{mathtools}
\usepackage{xcolor}

\DeclarePairedDelimiterX\set[1]\lbrace\rbrace{#1}

\usepackage{hyperref}
\hypersetup{colorlinks=true}
\usepackage[nameinlink]{cleveref}
\Crefname{figure}{Fig.}{Fig.}

\usepackage{amsfonts}
\usepackage{graphicx}
\usepackage{epstopdf}
\usepackage{algorithm}
\usepackage{algpseudocode}
\ifpdf
  \DeclareGraphicsExtensions{.eps,.pdf,.png,.jpg}
\else
  \DeclareGraphicsExtensions{.eps}
\fi
\renewcommand{\algorithmicrequire}{\textbf{Input:}}
\renewcommand{\algorithmicensure}{\textbf{Output:}}

\newcolumntype{L}{>{\centering\arraybackslash}m{6cm}}
\newcolumntype{M}{>{\centering\arraybackslash}m{11cm}}

\newtheorem*{FLPH}{Fundamental Lemma of Persistent Homology}
\newtheorem*{lemma*}{Lemma}
\newtheorem{lemma}{Lemma}
\newtheorem{proposition}{Proposition}
\newtheorem{theorem}{Theorem}
\theoremstyle{definition}
\newtheorem{definition}{Definition}
\newtheorem{remark}{Remark}
\newtheorem{example}{Example}
\newtheorem*{questions}{Questions}

\begin{document}
\title{Persistence Curves: A Canonical Framework for Summarizing Persistence Diagrams
}


\author{Yu-Min Chung}
\thanks{The majority of this work was done when Yu-Min Chung was employed at the Department of Mathematics and Statistics, University of North Carolina at Greensboro.}
\address{Eli Lilly and Company, Indianapolis, Indiana 46225, USA}
\email{yumchung@alumni.iu.edu}           

\author{Austin Lawson}
\address{Program of Informatics and Analytics, University of North Carolina at Greensboro, Greensboro, North Carolina 27412, USA}
\email{azlawson@uncg.edu}

\maketitle

\begin{abstract}
 Persistence diagrams are one of the main tools in the field of Topological Data Analysis (TDA). They contain fruitful information about the shape of data. The use of machine learning algorithms on the space of persistence diagrams proves to be challenging as the space lacks an inner product.  For that reason, transforming these diagrams in a way that is compatible with machine learning is an important topic currently researched in TDA. In this paper, our main contribution consists of three components.  First, we develop a general and unifying framework of vectorizing diagrams that we call the \textit{Persistence Curves} (PCs), and show that several well-known summaries, such as Persistence Landscapes, fall under the PC framework. Second, we propose several new summaries based on PC framework and provide a theoretical foundation for their stability analysis. Finally, we apply proposed PCs to two applications---texture classification and determining the parameters of a discrete dynamical system; their performances are competitive with other TDA methods. \end{abstract}



\section{Introduction}\label{sec:introduction}

Topological data analysis (TDA) is a rising field in mathematics, statistics, and computer science.  The fundamental concept of TDA is to understand the shape of data.  Due to its effectiveness and novelty, TDA has been applied to different scientific disciplines, such as neuroscience \cite{bendich2016persistent}, medical biology \cite{li2015identification}, sensor networks \cite{de2007coverage}, social networks \cite{carstens2013persistent}, physics \cite{donato2016persistent}, nanotechnology \cite{nakamura2015persistent},  material science \cite{saadatfar2017pore}.  The development of \textit{persistent homology} (see e.g. \cite{frosini1992measuring,ferri1998point, edelsbrunner2000topological, zomorodian2005computing}) is one of the driving forces in TDA. Persistent homology extracts topological information from a dataset by tracking the changes in topological features over some varying parameter. Information about those changes is stored in the \textit{persistence diagram}. 

Making statistical inferences or extracting meaningful information from persistence diagrams is an essential task in TDA.  
One may equip the set of persistence diagrams with a metric, namely the bottleneck distance or $p$-Wasserstein metric, and transform the set into a metric space \cite{cohen2007stability,mileyko2011probability}. As a metric space, one can apply distance-based machine learning algorithms such as $k$-nearest neighbors on the space of diagrams; however, many more machine learning algorithms require inputs coming from a Hilbert space. It has been shown that the space of persistence diagrams is not a Hilbert space \cite{mileyko2011probability,Bubenik_2018}. Furthermore, recent research has shown that the metric space of persistence diagrams is large \cite{bell2019space} and cannot embed into a Hilbert space \cite{bubenik2019embeddings, turner2019different,Carrire2019OnTM} with respect to the aforementioned metrics. Hence, to succeed in using modern machine learning algorithms on persistence diagrams, we must map them into a Hilbert space in some meaningful way, which has become a major research area in TDA.  This mapping process is also known as summarizing the persistence diagrams.

Kernel functions and vectorization of persistence diagrams are two main approaches to summarize persistence diagrams in a way that is compatible with non-distance based machine learning algorithms.  In the former, one constructs a kernel function, or a rule for quantifying the similarity of two persistence diagrams. This kernel function is then used in kernel-based machine learning algorithms such as support vector machines. This approach has been explored through in a bag-of-words approach \cite{li2014persistence}, kernel SVM for persistence \cite{reininghaus2015stable}, persistence intensity functions \cite{chen2015statistical}, and persistence weighted Gaussian kernel \cite{kusano2016persistence}. The vectorization of persistence diagrams has proven quite popular in recent literature and we can find this summarization type in the form of persistence landscapes~\cite{bubenik2015statistical}, persistence images~\cite{adams2017persistence}, persistence indicator functions~\cite{rieck2019topological}, general functional summaries \cite{berry2018functional}, persistent entropy~\cite{persistentEntropyStability}, and the Euler Characteristic Curve~\cite{richardson2014efficient}.  
Indeed, several of these vectorization methods share some properties.

Our main contribution, called \textit{persistence curves} framework, is a general unifying framework that many vectorizations fall under. This allows for a stronger theoretical analysis of diagram vectorizations. It is canonical as it generalizes the idea of Fundamental Lemma of Persistent Homology \cite{edelsbrunner2010computational}, making it an intuitive framework. It is flexible and interpretable as one may design new summaries based on different situations or applications. It is capable of generating competitively applicable summaries at a much lower computational cost than other examples in the literature. 
These advantages of persistence curves framework echo those qualities of a good vectorization method outlined in \cite{adams2017persistence} which are the following: 
\begin{itemize}
  \setlength{\itemsep}{-1pt}
    \item[Quality 1] The output of the representation is a vector in $\mathbb{R}^n$.
    \item[Quality 2] The representation is stable with respect to the input noise.
    \item[Quality 3] The representation is efficient to compute.
    \item[Quality 4] The representation maintains an interpretable connection to the original persistence diagram.
    \item[Quality 5] The representation allows one to adjust the relative importance of points in different regions of the persistence diagram.
\end{itemize}
We will show throughout this paper that our proposed framework can generate vectorizations that possess these qualities. The outline of the paper is as follows.

In \Cref{sec:background}, we provide a brief introduction to persistent homology, persistence diagrams, the bottleneck distance, and the Wasserstein distances. More importantly, we review the Fundamental Lemma of Persistent Homology, that serves as the inspiration for our main construction.  In \Cref{sec:PersistenceCurves}, we propose the persistence curve (PC) framework, which addresses Quality 1. We use the framework to propose new interpretable persistence diagram summaries (Quality 4 and 5) and show the framework can generate other popular summaries in the literature.  In \Cref{sec:a general bound}, we provide an analysis of a specific class of PCs. \Cref{thm:main stability thm}, our main theorem, provides a general bound with respect to the bottleneck and $1$-Wasserstein distances. We apply \Cref{thm:main stability thm} in \Cref{sec:application of main thm} to produce explicit bounds for some selected curves. 
In \Cref{sec:computation and applications}, we address Quality 3 and show that our proposed summaries are much more efficient to compute than some other popular TDA methods. We also provide two applications of the curves proposed in \Cref{sec:PersistenceCurves}.
In one application, we use the new summaries to classify images of textures from the UIUCTex, KTH-TIPS2b, and Outex databases. In the other application we utilize these summaries to determine a parameter in a discrete dynamical system. In both applications, we compare our proposed curves to other TDA methods including persistence landscapes, persistent entropy, and persistence images. Then we provide an experiment to further address Quality 2 by adding noise to the KTH-TIPS2b and Outex databases. Finally we discuss some limitations and considerations of the framework. We conclude this paper in \Cref{sec:conclusion} by providing several directions for future research and applications.

\section{Background}
\label{sec:background}

Homology provides a discrete object as a descriptor of a topological space that is invariant under continuous deformations. The $k$-th homology group $H_k(X)$ of a space $X$ is often used to calculate the \textbf{{$k$-th Betti}} number, denoted by $\beta_k(X)$, which counts the number of $k$-dimensional holes of $X$ (see e.g. \cite{kaczynski2004computational,rotman1998introduction} for more details about homology). 
A \textbf{filtration} of $X$ is an increasing sequence of subspaces  of $X$, $\emptyset = X_0\subset X_1\subset\ldots\subset X_n = X$. Informally, persistent homology tracks the appearance (birth) and disappearance (death) of homological features over a filtration of $X$. More details about persistent homology can be found e.g. in \cite{edelsbrunner2010computational}.
We collect the birth-death information and store it in a multi-set called a \textbf{persistence diagram}.
Persistence diagrams arise naturally through persistent homology; however, it is also helpful to define a persistence diagram in a more general way.

A \textbf{multi-set} is a collection of objects that are allowed to repeat. The number of times an object $s$ repeats is called its multiplicity and is denoted by $m(s)$. To distinguish a set from a multi-set, we follow the notation of  \cite[Section 1.2.4, p.~29]{hein2003discrete} and use square brackets $[~]$ to denote a multi-set. Define the diagonal multi-set $\Delta = [(x,x)\mid x\in\mathbb{R}, ~m(x,x) = \infty]$. A \textbf{persistence diagram} $D$ is a union of two multi-sets \[D = \Delta \cup [(b,d)\mid b<d\in \mathbb{R}\cup\{\infty\},~ m(b,d)<\infty].\]
We call $D\setminus\Delta$ the \textbf{off-diagonal} of $D$. We follow the convention in TDA to assume that $m(x,x) = \infty$ in order to define \eqref{equ:Winfty distance} and \eqref{equ:Wp distance}.
In this paper, we assume that the off-diagonal of any diagram has finite cardinality including all multiplicities. We denote the set of all persistence diagrams as $\mathcal{D}$.

$\mathcal{D}$ is a metric space \cite{cohen2007stability, cohen2010lipschitz} when equipped with the bottleneck or $p$-Wasserstein distance, which we define below. 
For any $C$, $D \in \mathcal{D}$, the \textbf{bottleneck distance} is defined as
\begin{equation}
    \label{equ:Winfty distance}
W_\infty(C,D) = \inf\limits_{\stackrel{\hbox{bijections }}{{\eta:C\to D}}}\sup_{(b,d)\in C}\|(b,d)-\eta(b,d)\|_\infty,
\end{equation}
and the \textbf{p-Wasserstein distance} is defined as
\begin{equation}
    \label{equ:Wp distance}
W_p(C,D) =\inf\limits_{\stackrel{\hbox{bijections }}{{\eta:C\to D}}}\left(\sum_{(b,d)\in C}\|(b,d)-\eta(b,d)\|_\infty^p\right)^{\frac{1}{p}}.\end{equation}
In this work, we focus on the $W_\infty$ and $W_1$ distances, both of which have stability results \cite{cohen2007stability, mileyko2011probability} that make them viable options for describing the distance between two diagrams. 

The Fundamental Lemma of Persistent Homology (FLPH) \cite[p.~118]{edelsbrunner2010computational} bridges the gap between the homology of the spaces in a filtration and persistent homology associated to that filtration. Define the \textbf{fundamental box} at $t\in\mathbb{R}$ as the multi-set $F_t = [(x,y)\mid x\le t < y\in\mathbb{R}, m(x,y) = \infty]$.  For a persistence diagram $D$, we define $D_t = F_t\cap D$. Note that $m(x,y)=\infty$ in the definition of $F_t$ is to ensure enough points for $D_t$.

\begin{figure}
\centering
\begin{subfigure}{0.15\linewidth}
\centering
\includegraphics[width=\linewidth]{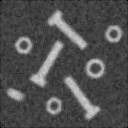}
\caption{Original image $X$}
\end{subfigure}
~
\begin{subfigure}{0.15\linewidth}
\centering
\includegraphics[width=\linewidth]{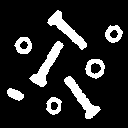}
\caption{Threshold at $110$}
\end{subfigure}~
\begin{subfigure}{0.33\linewidth}
\centering
\includegraphics[width=\linewidth]{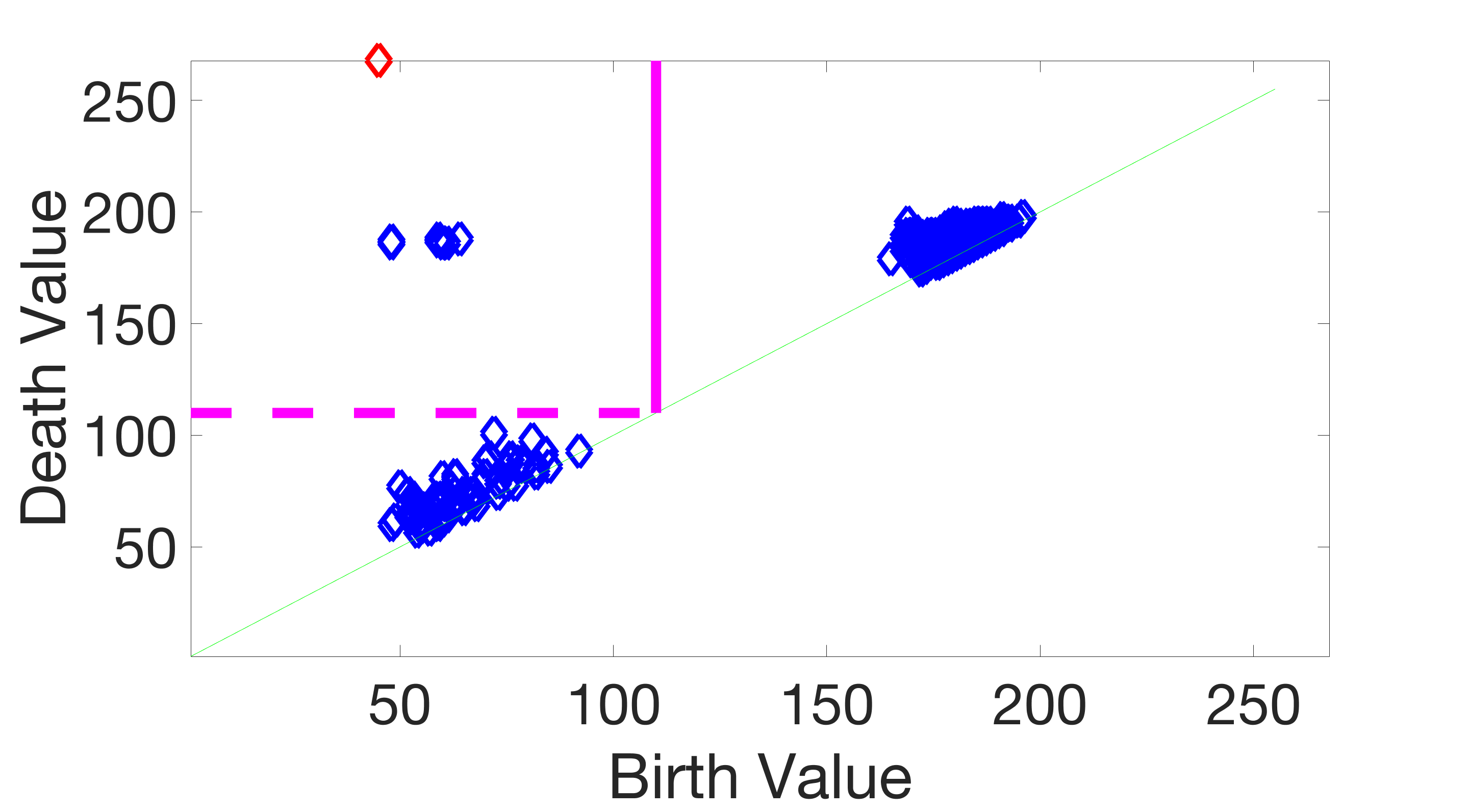}
\caption{0-dimensional diagram}
\end{subfigure}
\begin{subfigure}{0.33\linewidth}
\centering
\includegraphics[width=\linewidth]{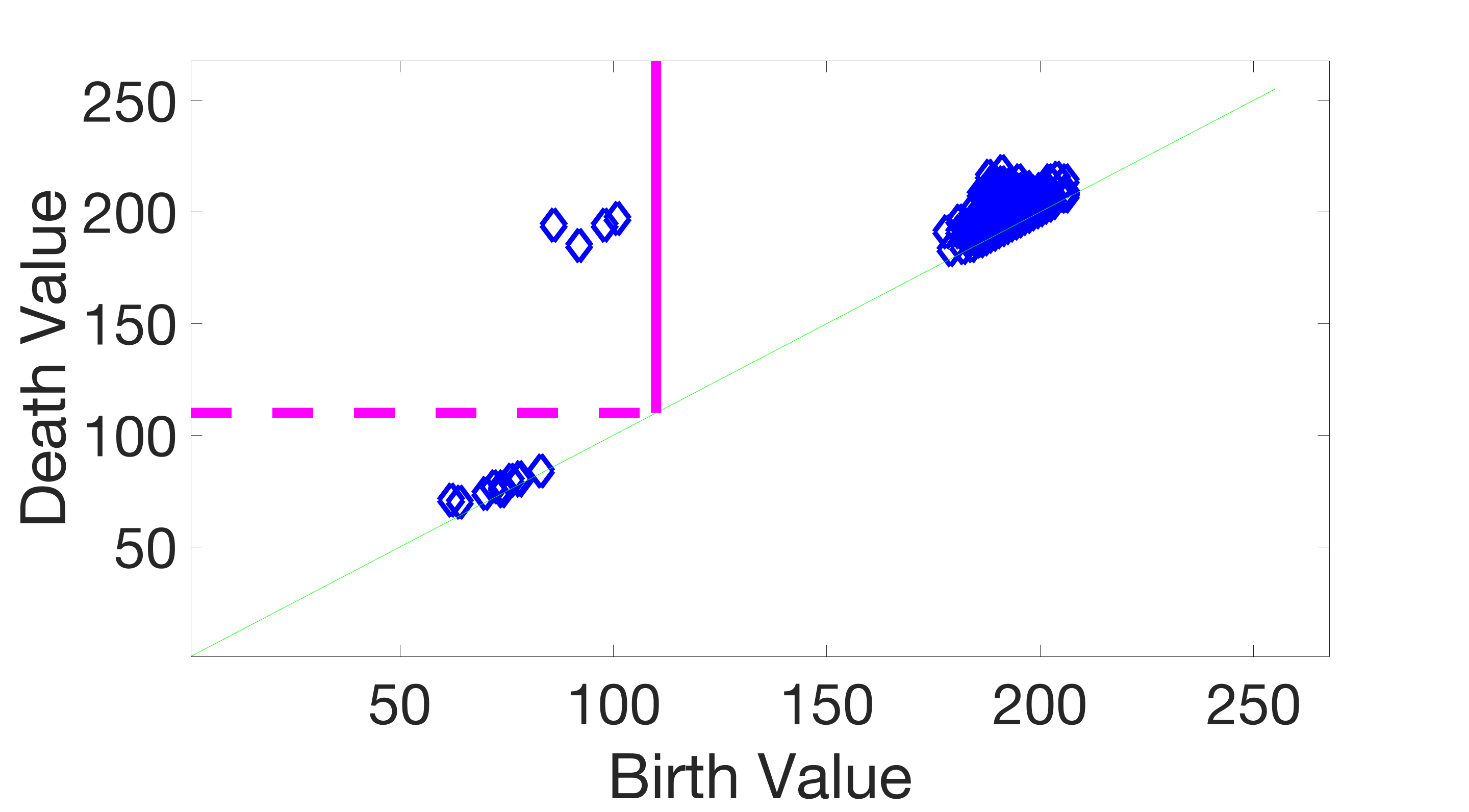}

\caption{1-dimensional diagram}
\end{subfigure}
\caption{An illustration of persistence diagrams of a grayscale image via sublevel set filtration.  (a) By visual inspection, one expects the Betti numbers are $(8,4)$.  (b) A binary image is obtained from thresholding the $X$ at $110$. The Betti numbers of this binary image are $(8,4)$.  (c)-(d) The 0 and 1-dimensional persistence diagrams of $X$.  The region enclosed by the pink dotted line ($d>110$) and solid line ($b\leq 110$) represents the multi-set $D_{110} = [ (b,d)|~b\leq 110,~d>110]$.  $\beta_0(X_{110})=8$ and $\beta_1(X_{110})=4$.}
\label{fig:persistent homology parts example}
\end{figure}

\begin{FLPH}
Let $D$ be the $k$-dimensional diagram with respect to a filtration $\{ X_i\}_i$. Then
\begin{equation}
    \beta_k(X_t) = \#[(b,d)\in D\mid b\le t < d~] = \#( F_t\cap D) = \#D_t.
    \label{equ:fundamental box betti}
\end{equation}
\end{FLPH}
In other words, FLPH states that the $k$-th Betti number of the space at filtration value $t$ can be found by counting the points of the persistence diagram (with multiplicity) that lie in fundamental box. \Cref{fig:persistent homology parts example} demonstrates FLPH on a simple 8-bit grayscale image. 
FLPH is the inspiration for the PC Framework. Our construction extends \eqref{equ:fundamental box betti} to a more general format. We provide details of PC Framework derivation in the next section. 

\section{The Persistence Curve Framework}
\label{sec:PersistenceCurves}
As we discussed in \Cref{sec:introduction}, summarizing persistence diagrams is a major research area in TDA. 
The framework we propose in this paper generates functional summaries of persistence diagrams that are compatible with statistical learning methods and retain topological information. To motivate the PC framework, we abstract FLPH as follows:
\begin{equation}
\label{equ:derivation of pc}
    \beta(t) = \# D_t = \sum_{(b,d)\in D_t} 1 = \sum_{(b,d)\in D_t} \psi(b,d,t) = \underbrace{\sum}_{{\text{(III)}}} \underbrace{[~ \overbrace{\psi(b,d,t)}^{\text{(I)}}~|~(b,d)\in D_t~ ] }_{\text{(II)}},
\end{equation}
{where $\psi(b,d,t) = 1$ if $b< d$ and $0$ otherwise. The right side of \eqref{equ:derivation of pc} can be considered as a general form from which we can generate new summaries by considering its three major components, labelled as (I), (II), and (III).  For (I), we can take $\psi$ is a user-defined function, and it can be chosen to fit the situation.  In (II), for a given $t$, we evaluate the function $\psi$ at all elements of $D_t$. 
Finally, in (III), we select an operator that maps multi-sets to real numbers, which summarizes the function values obtained in (II). In this case, we use summation, and in general, this operator is also user-defined. We elaborate on this in the formal definition of the PC Framework below. }

Recall that $\mathcal{D}$ denotes the set of all persistence diagrams. Let $\mathcal{F}$ be the set of all functions $\psi:\mathcal{D}\times \mathbb{R}^3\to\mathbb{R}$ with $\psi(D;b,b,t) = 0$ for all $(b,b) \in \Delta$. To ease the notation, we refer to $\psi(D;b,d,t)$ as $\psi(b,d,t)$ when $D$ is understood.
Moreover, when $\psi$ does not depend on $t$, we write $\psi(b,d)$.
Let $\mathcal{T}$ be the set of \textbf{summary statistics} or operators that map multi-sets to real numbers (e.g., sum, max, median, etc.).  Finally, let $\mathcal{R}$ represent the set of functions on $\mathbb{R}$. 
Our definition of the PC Framework presented below follows the definition proposed in Lawson's dissertation \cite{lawson2019preservation}.
\begin{definition}\cite[Def III.2 pg. 51]{lawson2019preservation}
\label{def:persistence curve}
We define a map $P:\mathcal{D}\times\mathcal{F}\times\mathcal{T}\to\mathcal{R}$ where  
\[P(D,\psi, T)(t) = T([\psi(D;b,d,t)\mid (b,d)\in D_t]),~t\in\mathbb{R}.\] 
The function $P(D,\psi, T)$ is called the \textbf{persistence curve} of $D$ with respect to $\psi$ and $T$.
\end{definition}

\begin{remark}
By the time of the submission, it came to our attention that the main idea of PCs in \Cref{def:persistence curve}
presented here is nearly equivalent to that of the PersLay construction introduced by Carri\`{e}re et. al. \cite{carrire2019perslay}. One of the main differences is that theoretically, PCs are functions whose values are calculated based on the fundamental box whereas PersLay produces vectors whose input values have no required structure. {Moreover, as currently defined, PersLay cannot produce curves equivalent to our Normalized and Entropy, which we define in \Cref{sec:application of main thm}.}
To the best of our knowledge, the definition of PCs first appeared in the work by Chung et. al. 2018 \cite{ChungHuLawsonSmyth} and were formally introduced by Lawson in his doctoral dissertation \cite{lawson2019preservation}.  This paper is a much more in-depth study of Lawson's thesis work. Specifically, we expand on the ideas, provide details of stability analysis, and conduct extensive numerical experiments.

\end{remark}

\begin{remark}
We remark on methods for handling infinite generators whose death values is by convention $\infty$. If there is a global maximum finite death value for the space, one may set all infinite death values to this maximum. 
If no global maximum death exists one can choose to replace the infinite generator with the max death value in a given diagram. Whether or not a global maximum death time exists, one can also choose to neglect infinite generators.
\end{remark}

\begin{remark}
For the purpose of creating new summaries, there are no regularity conditions on $\psi$ in \Cref{def:persistence curve}. However, the stability of a PC as a function of diagrams may require additional conditions on $\psi$ as we will discuss in \Cref{sec:a general bound}. 
\end{remark}

\Cref{def:persistence curve} is in a general format.  To further characterize it, with a mild assumption on the operator $T$, one may rewrite a persistence in terms of indicator functions.  This alternate view can also be seen as functional defined on the persistence barcodes, such as \cite{persistentEntropyStability}.
\begin{proposition}
\label{prop:pc realization}
Let $\psi$ be as in \Cref{def:persistence curve} and $[0]$ be a multi-set with $m(0)>0$, and $A\subset [(x,y)\mid x<y\in\mathbb{R}\cup\{\infty\}]$. Assume that the operator $T$ in \Cref{def:persistence curve} satisfies $T(\psi(A)) = T(\psi(A)\cup[0])$. Then
\begin{align}
P(D,\psi,T)(t) = T\left([\psi(b,d,t)\mid (b,d)\in D_t]\right) = T\left( [\psi(b,d,t)\chi_{[b,d)}(t)\mid (b,d)\in D] \right),
\label{equ:pc realization}
\end{align}
where $\chi_{[b,d)}$ is the indicator function on the interval $[b,d)$, i.e. $\chi_{[b,d)}(t) = 1$ if $t\in [b,d)$ and 0 otherwise.
\end{proposition}
\begin{proof}
Given any $t$, the persistence diagram, $D$, can be decomposed as $D= D_t \cup D_t^C$.  Moreover, by the definition of indicator functions, it is straightforward to verify that if $(b,d)\in D_t$, $\chi_{[b,d)}(t) = 1$, and if $(b,d)\notin D_t$, $\chi_{[b,d)}(t) = 0$.   Thus, we have
\begin{align*}
    [\psi(b,d,t)\chi_{[b,d)}(t)\mid (b,d)\in D]&=[\psi(b,d,t)\chi_{[b,d)}(t)\mid (b,d)\in D_t] \cup [\psi(b,d,t)\chi_{[b,d)}(t)\mid (b,d) \notin D_t]\\
    &=[\psi(b,d,t)\mid (b,d)\in D_t] \cup [0]
\end{align*}
Finally, we apply the operator $T$ on both sides and use the assumption $T(\psi(A)) = T(\psi(A)\cup[0])$ to obtain
\begin{align*}
    T\left( [\psi(b,d,t)\chi_{[b,d)}(t)\mid (b,d)\in D] \right)  &= T\left([\psi(b,d,t)\mid (b,d)\in D_t]\cup [0]\right)\\ &= T\left([\psi(b,d,t)\mid (b,d)\in D_t]\right) \\ &= P(D,\psi,T)(t).
\end{align*}
This completes the proof of \Cref{prop:pc realization}.
\end{proof}

 We call \eqref{equ:pc realization} the \textbf{indicator function realization}\footnote{Alternatively, if a persistence curve has an indicator realization, we can also consider it as a linear representation as seen in \cite{divol2020understanding}. Specifically, suppose $(x,y)\in\Omega:=\{(x,y)\in\mathbb{R}\mid x\le y\}$. We can define a function $f:(x,y)\in\Omega\mapsto(t\mapsto\psi(x,y,t)\chi_{[b,d)}(t))$. Then, by mapping a diagram $D$ as a measure $\mu_D = \sum_{(b,d)\in D}\delta(b,d)$ where $\delta$ is the Dirac measure. We then arrive at the linear representation $D\mapsto\mu_D(\psi):=\int \psi\mathrm{d}\mu$.} of the PC. We find that in practice, \Cref{def:persistence curve} is suitable for the numerical implementation while the indicator function realization \eqref{equ:pc realization} (when applicable) is suitable and sometimes preferable for theoretic analysis.  In the following example, we create a PC based on \Cref{def:persistence curve} and demonstrate both \Cref{def:persistence curve} and its indicator function realization.  

\begin{example}
\label{ex:lifespan curve}
Let $\psi(b,d,t) = d-b$ and $T = \Sigma$.  Then by \eqref{equ:pc realization},
\begin{equation*}
    \mathbf{l}(D)(t):=
    P(D, \psi, \Sigma)(t) = \sum ([ d-b\mid (b,d)\in D_t]),
\end{equation*}
which we will refer as the \textbf{lifespan curve}.
We evaluate $\mathbf{l}(D)(2)$ in two different ways given that $D = \{ (1,2), (2,4), (2,4), (3,5), (1,5)\}$.  By \Cref{def:persistence curve}, we first find $D_2 = [~(b,d)\in D \mid~ b\le 2<d ] = [(2,4), (2,4), (1,5)]$ and then, $\mathbf{l}(D)(2) = \sum~ [~ d-b \mid~ (b,d)\in D_2 ] = (4-2) + (4-2) + (5-1) = 8$.   On the other hand, since $T = \sum$, by \eqref{equ:pc realization}, we obtain that $\mathbf{l}(D)(t) = \chi_{[1,2)} (t) + 2\chi_{[2,4)}(t) + 2\chi_{[2,4)}(t) + 2\chi_{[3,5)}(t) + 4\chi_{[1,5)}(t)$. 
Thus, $\mathbf{l}(D)(2) = \chi_{[1,2)} (2) + 2\chi_{[2,4)}(2) + 2\chi_{[2,4)}(2) + 2\chi_{[3,5)}(2) + 4\chi_{[1,5)}(2) = 8.$
\end{example}

\begin{remark}
{We remark on the condition on $T$ in \Cref{prop:pc realization}.  Not all of operation would satisfy the condition.}
For instance, let $T = \text{Avg}$, be the average operator and $\psi(b,d,t) = d-b$. Then for a given $t$, the resulting $P(D,\psi,\text{Avg})$ computes the average lifespan 
in $D_t$.  Let $D$ be as in \Cref{ex:lifespan curve}. We compute $P(D,\psi,\text{Avg})$ at $t=2$ by \Cref{def:persistence curve}. First, we note that $D_2=[(2,4),~(2,4),~(1,5)]$. Next we consider the multi-set $\psi(D_2) = [2,~2,~4]$. Finally, we take the average so that $P(D,\psi,\text{Avg})(2) = \text{Avg}[2,~2,~4] = \frac{8}{3}$. On the other hand, if we attempt the indicator realization realization~\eqref{equ:pc realization}, then
\[
\text{Avg}~[\psi(b,d,t)\chi_{[b,d)}(t)\mid (b,d)\in D] = \text{Avg}~[0,~2,~2,~0,~4] = \frac{8}{5} \neq \frac{8}{3}.
\]
Thus, when $T = \text{Avg}$, the example shows that the indicator realization fails.
\end{remark}

 The lifespan curve tracks lifespan information over the filtration. One can think of it as a topological intensity function that accounts for the size or intensity of topological features. This curve is one of many examples of PCs immediately generated by common diagram statistics. The following example can be found in \cite{persistentEntropyStability}, where the concept of entropy introduced to TDA.

\begin{example}
\label{ex:persistent entropy curve}
In \cite{persistentEntropyStability} a summary function based on persistent entropy was defined as: \[S(D)(t) = -\sum w(t) \frac{d-b}{L^D}\log\left(\frac{d-b}{L^D}\right),\] where $L^D = \sum_{(b,d)\in D}(d-b)$ and $w(t) = 1$ if $b\leq t\leq d$ and $w(t)=0$ otherwise. We note that in this paper, we use $\log$ to represent the natural log.
Let $\psi = -\frac{d-b}{L}\log\frac{d-b}{L}$, and $T = \Sigma$.  We see that $\mathbf{le}(D) := P(D, \psi, T)$ is similar to $S(D)$. The difference is subtle, but ultimately $S(D)$ is defined using closed intervals $[b,d]$ while $\mathbf{le}(D)$ uses $[b,d)$. This difference is almost negligible and $\mathbf{le}(D)$ will enjoy the same stability as $S(D)$. In this work we refer to $\mathbf{le}$ as the \textbf{life entropy curve}.
\end{example}

\begin{example}
\label{ex:pd thresholding} The PD Thresholding method \cite{chung2018topological} was developed for image processing as a way to compute the optimal threshold value for an image.  The main idea is to define an objective function, and the optimal threshold will be chosen as the maximum of the objective function.  One major component of the objective function in \cite{chung2018topological} is $O(t) =\frac{1}{\# D_t} \sum_{\substack{ (b,d)\in D_t }} (d - t) (t - b)$.  The function $O(t)$ can be viewed as a PC if one lets $\psi = (d-t)(t-b)$ and $T$ be the average operator.
\end{example}
In the last two examples, we recognize persistence landscapes and persistence silhouettes as special cases of PCs. 
\begin{example}
\label{ex:persistence landscapes}
Let $\max_k(S)$ represent the $k$-th largest number of a set $S$.  Given a persistence diagram $D$, define
\[\Lambda_{(b,d)}(t) = 
\begin{cases}
0 & \hbox{if } t\notin [b,d]\\
t-b & \hbox{if } t\in[b,\frac{b+d}{2}]\\
d-t & \hbox{if } t\in(\frac{b+d}{2}, d]
\end{cases}.\]
Then the $k$-th Persistence Landscape \cite{bubenik2015statistical} is defined by $\lambda_k(t) = \max_k\{\Lambda_{(b,d)}(t)\mid (b,d)\in D\}$.  One can verify that $\Lambda_{(b,d)}(t) = \min\{t-b,d-t\}\chi_{[b,d)}$.  Thus, if $\psi(b,d,t) = \min\{t-b,d-t\}$ and $T = \max_k$, then $P(D,\psi,T) \equiv \lambda_k$. 

\end{example}

\begin{example}
Let $T=\Sigma$ and define $\psi(D,b,d,t) = \omega(D,b,d,t)\Lambda_{(b,d)}(t)$, where $\omega(D,b,d,t)$ is an arbitrary weight function, we recover the persistence silhouette function as defined in \cite{chazal2014stochastic}. For example, if $\omega(D;b,d,t) = \frac{(d-b)^p}{\sum_{(b,d)\in D}(d-b)^p}$, the we obtain the power-weighted silhouette.
\end{example}

\begin{remark}
The Euler Characteristic is a classic descriptor of topological spaces that pre-dates persistent homology. It is calculated by taking the alternating series of the space's Betti numbers. That is, for a topological space $X$, the Euler characteristic of $X$, is $EC(X) = \sum_{n=0}^\infty (-1)^n\beta_n(X)$. By applying the Euler Characteristic to spaces in a filtration of topological spaces, we recover the so-called Euler Characteristic Curve \cite{richardson2014efficient}. While it is not an example of a PC, it is derivable from the framework. Specifically, if $D^0,D^1,\ldots D^k\ldots$ are $k$-dimensional persistence diagrams arising from the same filtration then the Euler Characteristic Curve can be defined as the alternating sum of the Betti number curves, $ECC\equiv \sum_{n=0}^\infty (-1)^n \beta_n$. 
\end{remark}

Examples in this section demonstrate the PC framework for summarizing persistence diagrams and show that the framework subsumes many existing summaries. \Cref{tab:PC} displays several new summaries based on the PC framework and their stability results (that will be discussed in \Cref{sec:application of main thm}).  Among those curves we propose in \Cref{tab:PC}, there are three categories. In the first category, $\psi$ utilizes basic information from persistence diagrams, including the Betti ($\beta$), lifespan ($\mathbf{l}$) and midlife span ($\mathbf{ml}$), so we refer to those curves as basic PCs.  The second category, called normalized PCs, modifies basic curves by normalizing the function $\psi$. These normalized curves provide both theoretical and practical advantages as we will discuss in \Cref{sec:application of main thm} and \Cref{sec:computation and applications}. The {\bf s$\boldsymbol{\beta}$, sl},and \textbf{sml} curves are normalized versions of Betti, lifespan, and midlife curve, respectively, and they are new curves.  The third category, motivated by \cite{persistentEntropyStability}, is entropy-based PCs (that we will define in \Cref{def:entropy-based persistence curve}).  Both {\bf mle} and {\bf $\boldsymbol{\beta}$e} are new entropy-based functions using the midlife statistic and Betti number. {We remark that even though many of these summaries do not enjoy stability (e.g. Betti curve, ECC, etc.) they still often find their use and appear in the literature \cite{richardson2014efficient, umeda2017time} and modern TDA packages \cite{gudhiPersistenceRepresentations, tauzin2020giottotda}}.

In this section we have seen that the PC framework generates real-valued functions that carry interpretable information about the diagram (Quality 4). The choice of $\psi$ allows one to adjust the relative importance of points in the diagram (Quality 5). As we will see in \Cref{sec:computation and applications}, these real-valued functions are transformed into vectors in $\mathbb{R}^n$ (Quality 1).

\begin{table}
	\renewcommand{\arraystretch}{1.5}
	\begin{center}
{\scalebox{0.9}{
		\begin{tabular}{|c|c|c|c|c|c|} 
			\hline
			{\bf Name} & {\bf Notation} & { $\psi(b,d,t)$} & {T} & $W_{\infty}$ & $W_{1}$  \\\hline
			\multicolumn{6}{|l|}{
			\textit{Existing PCs}}\\\hline
			Betti number  &  $\boldsymbol{\beta}$ &$\chi_{b<d}(b,d)$ & sum & \xmark, \eqref{equ:betti Winfty bound}  & \xmark, \eqref{equ:betti W1 bound}  \\ \hline
			Life Entropy \cite{persistentEntropyStability}& {\bf le}&$ \displaystyle-\frac{d-b}{\sum(d-b)}\log\frac{d-b}{\sum(d-b)}$ &sum  & $\boldsymbol\bigcirc$ & $\boldsymbol\bigcirc$  \\ \hline
			PD Thresholding \cite{chung2018topological}& {$O$}&$ (d-t)(t-b)$ & avg  & $\boldsymbol{-}$ & $\boldsymbol{-}$ \\ \hline
			$k$-th Landscape \cite{bubenik2015statistical}& $\boldsymbol{\lambda_k}$ & $ \min\{t-b,d-t\}$&$\max_k$  & \cmark & \cmark  \\ \hline
			\multicolumn{6}{|l|}{
			\textit{PCs proposed in this work}} \\\hline
			Normalized Betti & $\mathbf{s}\boldsymbol{\beta}$ & $\frac{1}{n}\chi_{b<d}(b,d)$, $n=\# D$  & sum & \xmark, \eqref{equ:normalized betti Winfty bound} & \xmark, \eqref{equ:normalized betti W1 bound} \\ \hline
			Betti Entropy & $\boldsymbol{\beta}\mathbf{e}$ & $-\frac{1}{n}\log\frac{1}{n}\chi_{b<d}(b,d)$  & sum & \xmark, \eqref{equ:betti entropy Winfty bound} & \xmark, \eqref{equ:betti entropy W1 bound} \\ \hline
			Life & $\mathbf{l}$ & $d-b$  & sum & \xmark, \eqref{equ:lifespan Winfty bound} & $\boldsymbol\bigcirc$, \eqref{equ:lifespan W1 bound} \\ \hline
			Normalized Life & $\mathbf{sl}$ & $\displaystyle\frac{d-b}{\sum(d-b)}$  & sum & $\boldsymbol\bigcirc$, \eqref{equ:normalized life Winfty bound} & \cmark, \eqref{equ:normalized life W1 bound} \\ \hline
			Midlife & $\mathbf{ml}$ & $(b+d)/2$ & sum  & \xmark, \eqref{equ:midlife Winfty bound} & $\boldsymbol\bigcirc$, \eqref{equ:midlife W1 bound} \\ \hline
			Normalized Midlife & $\mathbf{sml}$& $(b+d)/\sum(d+b)$ & sum  & $\boldsymbol\bigcirc$,\eqref{equ:normalized midlife Winfty bound}  & \cmark, \eqref{equ:normalized midlife W1 bound} \\ \hline
			Midlife Entropy& $\mathbf{mle}$&$ \displaystyle-\frac{d+b}{\sum(d+b)}\log\frac{d+b}{\sum(d+b)}$  &sum & $\boldsymbol\bigcirc$, \eqref{equ:midlife entropy Winfty bound} & $\boldsymbol\bigcirc$, \eqref{equ:midlife entropy W1 bound} \\ \hline
		\end{tabular}
		}}
		\caption{Examples of PCs. In the top panel, existing summaries are realized in the PC framework. In the bottom panel, new summaries are proposed in this work.  
		The last two columns represent the stability with respect to the corresponding distance followed by an equation number for the corresponding bound.  ``\xmark'' means not stable; ``$\boldsymbol\bigcirc$'' means stable under additional assumptions; ``\cmark'' means stable; ``$\boldsymbol{-}$'' indicates that to the best of authors knowledge, the stability is unknown.  
		}
		\label{tab:PC}
	\end{center}
\end{table}

\section{Analysis on the PC Framework with $T=\Sigma$}
\label{sec:a general bound}
\bgroup
\def\arraystretch{1.5}
\begin{table}
    \centering
    \begin{tabular}{|c|l|}\hline
       {\bf Notation}&{\bf Description}\\\hline
       $\lor$ & max operator, i.e. $n_1 \lor n_2 = \max\{n_1, n_2\}$ \\ \hline
       $\land$ & min operator, i.e. $n_1 \land n_2 = \min\{n_1, n_2\}$ \\ \hline
        $ \mathcal{D}$ & space of persistence diagrams\\\hline        
        $ \mathcal{D}_{M,m,q}$ & $\{D\in \mathcal{D} ~|~ m\leq b < d \leq M,~d-b\geq q >0\}$\\\hline               
       $n^D$   & number of off-diagonal points in $D\in \mathcal{D}$ \\\hline
       $\eta$ & { A matching between $C,~D\in\mathcal{D}$, i.e. $x\in C \mapsto \eta(x)\in D$} \\ \hline
       $(\eta_b, \eta_d)$ & {$x = (b,d)\in C$ with corresponding points $\eta(x):= (\eta_{b},\eta_{d})\in D$} \\ \hline       
       $n_{\eta}$ & $\#\{(x,\eta(x))\mid x\in\Delta\implies\eta(x)\notin\Delta\}$
       \\ \hline       
         $ \displaystyle W_{\infty}(C,D)$ &  $\inf_{\eta:C\rightarrow D}\max_{1\leq i \leq n_{\eta}}\{|d_i - \eta_{d_i}| \lor |b_i - \eta_{b_i}|\} $ \\\hline
         $ \displaystyle W_{1}(C,D)$ &  $\inf_{\eta:C\rightarrow D}\sum_{i=1}^{n_{\eta}}\{|d_i - \eta_{d_i}| \lor |b_i - \eta_{b_i}|\} $ \\\hline
        $\displaystyle L^D $,  $\displaystyle L_{\infty}^D $ &  $\displaystyle\sum_{(b,d)\in D} ({d} - {b})$, $\displaystyle\max_{(b,d)\in D} ({d} - {b})$ \\\hline
        $\displaystyle \kappa_{1}(\psi, C, D) $ &  $\displaystyle\sum_{i=1}^{n^C}\max_{ t\in [b_i, d_i]} |\psi(b_i, d_i, t)| +  \sum_{j=1}^{n^D}\max_{t\in [{b_j}, {d_j}]} |\psi({b_j}, {d_j}, t)|$\\\hline
        $\displaystyle \kappa_{\infty}(\psi, C, D) $ &  $\displaystyle\max_{1\le i\le n^C}\max_{ t\in [b_i, d_i]} |\psi(b_i, d_i, t)| +  \max_{1\le j\le n^D}\max_{t\in [{b_j}, {d_j}]} |\psi({b_j}, {d_j}, t)|$\\\hline
        $\displaystyle\delta_{\infty}(\psi, C, D)$ &  $ \displaystyle\inf_{\eta:C\rightarrow D}\max_{\substack{1\leq i \leq n_{\eta} \\ t\in [b_i, d_i]\cap [\eta_{b_i}, \eta_{d_i}]}} |\psi(b_i, d_i, t) - \psi(\eta_{b_i}, \eta_{d_i}, t)|$ \\\hline
        $\displaystyle \delta_{1}(\psi, C, D)$ & $ \displaystyle\inf_{\eta:C\rightarrow D}\sum_{i=1}^{n_{\eta}}\max_{{ t\in [b_i, d_i]\cap [\eta_{b_i}, \eta_{d_i}]}} |\psi(b_i, d_i, t) - \psi(\eta_{b_i}, \eta_{d_i}, t)|$ \\\hline
    \end{tabular}
    \caption{The notation here pertains to two given diagrams $C$ and $D$. {Note that $\kappa_1$ and $\kappa_{\infty}$ are independent of the matching $\eta$. } }
    \label{tab:notations}
\end{table}
\egroup
In this section, we analyze some properties of PCs.  Specifically, given two persistence diagrams $C,~D\in\mathcal{D}$ and for fixed $\psi$ and $T=\Sigma$, we provide a bound for the difference (using the $L^1$-norm) between $P(C, \psi, T)$ and $P(D, \psi, T)$. Notations are summarized in \Cref{tab:notations}. 

Since we focus on $\Sigma$ operator, we can utilize the indicator functions realization \eqref{equ:pc realization}. Thus, investigating the difference between two PCs amounts to estimating the difference between two indicator functions.  Formally speaking, given two continuous and bounded functions $\psi_I$, $\psi_J$ on two finite intervals $I$, $J\subset\mathbb{R}$ respectively, we aim to estimate $\|\psi_I \chi_{I}-\psi_J\chi_J\|_1$. We see
\begin{align}
&\|\psi_I \chi_{I}-\psi_J\chi_J\|_1= \int_{I\setminus J} |\psi_I(t)|~dt + \int_{I\cap J} | \psi_I(t) - \psi_J(t)|~dt + \int_{J\setminus I} | \psi_J(t)| \nonumber \\
&\leq |I\setminus J|\cdot \max_{t\in I\setminus J} |\psi_I(t)| + |I\cap J|\cdot \max_{t\in I\cap J} |\psi_I(t) - \psi_J(t)| + |J\setminus I|\cdot \max_{t\in J\setminus I} |\psi_J(t)|, \label{equ:diff indi func real}
\end{align}
where $|I|$ is the length of the interval $I$.
In the case of \eqref{equ:pc realization}, each $I$ or $J$ is replaced by the half-open, half-closed interval $[b,d)\subset\mathbb{R}$ where $(b,d) \in D$ is a point on a persistence diagram.  In the following lemma, we provide detailed analysis in this content.
\begin{lemma}
\label{lem:estimate}
{Let $h(t) =  \psi_1(t) \chi_{[b_1, d_1)}(t) - \psi_2(t) \chi_{[b_2, d_2)}(t) $, where $b_1 \leq d_1$ and $b_2 \leq d_2$. We denote by $K = \max_{t\in [b_1,d_1]} |\psi_1(t)| + \max_{t\in [b_2,d_2]} |\psi_2(t)|$, and by $Q$ we denote the quantity $\max_{t\in[b_1,d_1]\cap[b_2,d_2]}|\psi_1(t)~-~\psi_2(t)|$. Suppose that $\psi_i:~[b_i,d_i] \rightarrow \mathbb{R}$ are continuous for $i=1,~2$. Then we have }
\begin{equation*}
 \| h \|_1 \leq K (|d_2 - d_1|\lor |b_2 - b_1|) + Q[(d_1 - b_1) \land (d_2-b_2)].
\end{equation*}
\end{lemma}
\begin{proof}
Without loss of generality, let $I = [b_1, d_1)$ and $J=[b_2, d_2)$.  There are three cases to consider.

\begin{tikzpicture}
\begin{axis}[
    title = {Case 1},
    axis y line=none,
    y=0.5cm/1.5,
    xtick={5, 9, 11, 13}, 
    xticklabels={$b_1$, $d_1$, $b_2$, $d_2$},
    restrict y to domain=0:1,
    axis lines=left,
    enlarge x limits=upper,
    scatter/classes={
        o={mark=*,fill=white}
    },
    scatter,
    scatter src=explicit symbolic,
    every axis plot post/.style={mark=*,thick},
    legend style={
        draw=none,
        at={(1,1)},
        anchor=south east
    },
    legend image post style={mark=none}
]
\addplot table[y expr=0,meta index=1, header=false] {
5 c
9 o
};
\addplot table [y expr=1,meta index=1, header=false] {
11 c
13 o
};
\end{axis}
\end{tikzpicture}

\noindent Case 1: $b_1\leq d_1 \leq b_2 \leq d_2$ and $I\cap J = \emptyset$. In this case, $I\setminus J = I$ and $J\setminus I = J$.  Thus, by \eqref{equ:diff indi func real}, and $d_1\leq b_2$, we obtain
\begin{align*}
\| h \|_1 &\le \int_{b_1}^{d_1} |\psi_1(t)| dt +\int_{b_2}^{d_2} |\psi_2(t)| dt \leq  \max_{t\in [b_1,d_1]} |\psi_1(t)| ~(d_1 - b_1) + \max_{t\in [b_2,d_2]} |\psi_2(t)| (d_2 - b_2)\\&\leq K(|d_1-b_1|\lor|d_2-b_2|)\underbrace{\leq}_{\because~d_1\leq b_2} K(|b_2-b_1|\lor|d_2-d_1|).
\end{align*}

\begin{tikzpicture}
\begin{axis}[
    title = {Case 2},
    axis y line=none,
    y=0.5cm/1.5,
    xtick={5, 7, 9, 13}, 
    xticklabels={$b_1$, $b_2$, $d_1$, $d_2$},
    restrict y to domain=0:1,
    axis lines=left,
    enlarge x limits=upper,
    scatter/classes={
        o={mark=*,fill=white}
    },
    scatter,
    scatter src=explicit symbolic,
    every axis plot post/.style={mark=*,thick},
    legend style={
        draw=none,
        at={(1,1)},
        anchor=south east
    },
    legend image post style={mark=none}
]
\addplot table [y expr=0,meta index=1, header=false] {
5 c
9 o
};
\addplot table [y expr=1,meta index=1, header=false] {
7 c
13 o
};
\end{axis}
\end{tikzpicture}

\noindent Case 2: $b_1 \leq b_2 \leq d_1 \leq d_2$ and $I\cap J = [b_2, d_1)$. In this case, $I\setminus J = [b_1, b_2)$ and $J\setminus I = [d_1, d_2)$.  By \eqref{equ:diff indi func real}, we obtain
\begin{align*}
    \| h \|_1 &\le \int_{b_1}^{b_2} |\psi_1(t)| dt +\int_{b_2}^{d_1} |\psi_1(t) - \psi_2(t)| dt +  \int_{d_1}^{d_2} |\psi_2(t)| dt \\
    &\leq K(|b_2- b_1|\lor|d_2-d_1|) + \underbrace{(d_1 - b_2)}_{\text{Next step.}}\max_{t\in [b_2, d_1]}|\psi_1(t) - \psi_2(t)|.
\end{align*}
It remains to show that $(d_1 - b_2) \leq (d_1 - b_1) \land (d_2 - b_2)$.  Since in this case $b_1\leq b_2$ and $d_1\leq d_2$, then $(d_1 - b_2) \leq (d_1 - b_1) $ and $(d_1 - b_2) \leq (d_2 - b_2)$.
Therefore, we have
\begin{align*}
   \| h \|_1 \leq  K(|b_2- b_1|\lor|d_2-d_1|) + (d_1 - b_1)\land (d_2-b_2) \max_{t\in[b_1,d_1]\cap[b_2,d_2]}|\psi_1(t) - \psi_2(t)|.
\end{align*}

\begin{tikzpicture}
\begin{axis}[
    title = {Case 3},
    axis y line=none,
    y=0.5cm/1.5,
    xtick={5, 7, 9, 13}, 
    xticklabels={$b_1$, $b_2$, $d_2$, $d_1$},
    restrict y to domain=0:1,
    axis lines=left,
    enlarge x limits=upper,
    scatter/classes={
        o={mark=*,fill=white}
    },
    scatter,
    scatter src=explicit symbolic,
    every axis plot post/.style={mark=*,thick},
    legend style={
        draw=none,
        at={(1,1)},
        anchor=south east
    },
    legend image post style={mark=none}
]
\addplot table [y expr=0,meta index=1, header=false] {
5 c
13 o
};
\addplot table [y expr=1,meta index=1, header=false] {
7 c
9 o
};
\end{axis}
\end{tikzpicture}

\noindent Case 3: $b_1 \leq b_2 \leq d_2 \leq d_1$  and $I\cap J = [b_2, d_2)$.  In this case, $I\setminus J = [b_1, b_2)$ and $J\setminus I =[d_2, d_1)$.
By \eqref{equ:diff indi func real} again, we obtain
\begin{align*}
    \| h \|_1 &= \int_{b_1}^{b_2} |\psi_1(t)| dt +\int_{b_2}^{d_2} |\psi_1(t) - \psi_2(t)| dt +  \int_{d_2}^{d_1} |\psi_1(t)| dt \\
    &\leq K( |b_2- b_1|\lor |d_2-d_1|) + (d_2 - b_2)\max_{t\in [b_2,d_2]}|\psi_1(t) - \psi_2(t)|.
\end{align*}
Similarly, in this case, it is straightforward to observe that $(d_2 - b_2) \leq (d_1-b_1) \land (d_2-b_2)$.  Hence, we obtain
\begin{align*}
   \| h \|_1 \leq  K (|b_2- b_1|\lor |d_2-d_1|) + (d_1 - b_1)\land (d_2-b_2) \max_{t\in[b_1,d_1]\cap[b_2,d_2]}|\psi_1(t) - \psi_2(t)|.
\end{align*}
Putting together Case 1, 2 and 3 completes the proof. 
\end{proof}

The bound in the \Cref{lem:estimate} consists of two terms $K (|b_2- b_1|\lor |d_2-d_1|)$ and $(d_1 - b_1)\land (d_2-b_2) \max_{t\in[b_1,d_1]\cap[b_2,d_2]}|\psi_1(t) - \psi_2(t)|$.  In the first term, $K$ is a constant based on the user-defined function $\psi$ and $(|b_2- b_1|\lor |d_2-d_1|)$ will contribute to the distance between two persistence diagrams.  In the second term, $(d_1 - b_1)\land (d_2-b_2)$ corresponds to the lifespan of the persistence diagram points.
\Cref{lem:estimate} is essential in proving our main theorem.
\begin{theorem}
\label{thm:main stability thm}
Let $C,~D\in \mathcal{D}$. Let $T$ be the $\Sigma$ operator.  Suppose that $T(\emptyset) = 0$.  We adopt the notations in \Cref{tab:notations}. Assume that for each $D\in\mathcal{D}$ and $(b,d)\in D$, $\psi(D;b,d,\cdot)$ is a continuous function. 
Then the following bounds hold
\begin{align}
    &\| P(C, \psi, \Sigma) - P(D, \psi, \Sigma)\|_1 \leq \kappa_{1}(\psi, C, D) W_{\infty}(C,D)+ (L^C \land L^D) \delta_{\infty}(\psi,C,D); \label{equ:main estimate} \\
    &\| P(C, \psi, \Sigma) - P(D, \psi, \Sigma)\|_1 \leq \kappa_{\infty}(\psi, C, D) W_{1}(C,D)+ (L_{\infty}^C \land L_{\infty}^D) \delta_{1}(\psi,C,D).     \label{equ:main estimate W1 dist}
\end{align}
\end{theorem}
\begin{proof}
{Let $\eta$ be a matching between $C$ and $D$. We need consider only the finite set $\{(x,\eta(x))\mid x\in\Delta\Rightarrow\eta(x)\notin\Delta\}$ with cardinality $n_\eta\le |C|+|D|$. We can order these pairs $x_i = (b_i,d_i)\in C$ with corresponding points $\eta(x_i) = (\eta_{b_i},\eta_{d_i})\in D$ for $i=1,\ldots n$.} Now, take the difference and consider the $1$-norm
\begin{align}
\| P(C, \psi, \Sigma) - P(D, \psi, \Sigma) \|_1   &= \left\| \sum_{i=1}^{n_{\eta}} ( \psi(C;b_i, d_i) \chi_{[b_i, d_i)} - \psi(D;\eta_{b_i}, \eta_{d_i}) \chi_{[\eta_{b_i}, \eta_{d_i})} \right\|_1  \nonumber \\
&\leq \sum_{i=1}^{n_{\eta}} \left\| \psi(C;b_i, d_i) \chi_{[b_i, d_i)} - \psi(D;\eta_{b_i}, \eta_{d_i}) \chi_{[\eta_{b_i}, \eta_{d_i})} \right\|_1. \label{proof:interstep}
\end{align}
To ease notation, define $\psi_i^C(t) = \psi(C;b_i, d_i, t)$ and $\psi_{\eta(i)}^D(t) = \psi(D;\eta_{b_i}, \eta_{d_i}, t)$. Then by \Cref{lem:estimate}, each norm in \eqref{proof:interstep} is dominated by
\begin{align}
&\left\| \psi_i^C \chi_{[b_i, d_i)} - \psi_{\eta(i)}^D\chi_{[\eta_{b_i}, \eta_{d_i})} \right\|_1 \nonumber
\leq \\& K_i  (\left|b_i - \eta_{b_i}\right|\lor \left|d_i-\eta_{d_i}\right|)+ \left[(d_i-b_i) \land (\eta_{d_i} - \eta_{b_i})\right]\max_{t\in [b_i, d_i]\cap [\eta_{b_i}, \eta_{d_i}]} |\psi_i^C(t) - \psi_{\eta(i)}^D(t)|, \label{equ:estimate each int}
\end{align}
where \[ K_i = \max_{ t\in [b_i, d_i]} \left|\psi_i^C(t)\right| + \max_{ t\in [\eta_{b_i}, \eta_{d_i}]} \left|\psi_{\eta(i)}^D(t)\right|.\]
Therefore, by \eqref{equ:estimate each int}, the inequality~\eqref{proof:interstep} becomes
\begin{align}
&\|P(C,\psi,\Sigma)-P(D,\psi,\Sigma)\|_1 \nonumber\\
& \leq \sum_{i=1}^{n_{\eta}} \bigg[ K_i  (|b_i - \eta_{b_i}|\lor |d_i-\eta_{d_i}|) +  [(d_i-b_i) \land (\eta_{d_i} - \eta_{b_i})] \max_{t\in [b_i, d_i]\cap [\eta_{b_i}, \eta_{d_i}]} |\psi_i^C(t) - \psi_{\eta(i)}^D(t)| \bigg]. \label{proof:branch}
\end{align}
There are multiple ways to estimate the inequality \eqref{proof:branch}.  In order to obtain~\eqref{equ:main estimate}, we first observe that { since by assumption $\psi(b,b,t)=0$, $\sum_{i=1}^{n_{\eta}} K_i = \kappa_{1}(\psi,C,D)$.  
}  Then, we perform the following steps to obtain
\begin{align*}
\eqref{proof:branch}&\leq  \max_{1\leq i \leq n_{\eta}}(|b_i - \eta_{b_i}|\lor |d_i-\eta_{d_i}|)~ \bigg(\sum_{i=1}^{n_{\eta}} K_i\bigg) +  \\ &\quad\quad\quad\max_{\substack{1\leq i \leq n_{\eta} \\ t\in [b_i, d_i]\cap [\eta_{b_i}, \eta_{d_i}]}} \left|\psi_i^C(t) - \psi_{\eta(i)}^D(t)\right| ~\bigg[\sum_{i=1}^{n_{\eta}} (d_i-b_i) \land (\eta_{d_i} - \eta_{b_i})\bigg] \\[10pt] 
\end{align*}
{
We take the infimum over $\eta$ to retrieve
\[\|P(C,\psi,\Sigma)-P(D,\psi,\Sigma)\|_1\le\kappa_{1}(\psi, C, D) W_{\infty}(C,D) +(L^C \land L^D) \delta_{\infty}(\psi, C, D).\]}
This completes the proof of \eqref{equ:main estimate}. 

The proof of \eqref{equ:main estimate W1 dist} is similar with a small alteration to \eqref{proof:branch}.
{
Note that because $\psi(b,b,t) = 0$, we have $\max_{1\le i \le n_{\eta}} K_i = \kappa_\infty(\psi,C,D)$,} where {$\kappa_{\infty}$ is independent of $\eta$,} and \\ $\left[\left(d_i-b_i\right) \land \left(\eta_{d_i} - \eta_{b_i}\right)\right]\le L^C_\infty\land L^D_\infty$, { which is also independent of $\eta$.} Thus,
\begin{align*}
    \eqref{proof:branch} &\le  \left(\max_{1\le i \le n_{\eta}} K_i\right) \sum_{i=1}^{n_{\eta}}\left(\left|b_i - \eta_{b_i}\right|\lor\left|d_i-\eta_{d_i}\right|\right) +  \\ & \quad\quad\quad (L^C_\infty\land L^D_\infty)\sum_{i=1}^{n_{\eta}}\max_{t\in [b_i, d_i]\cap [\eta_{b_i}, \eta_{d_i}]} \left|\psi_i^C(t) - \psi_{\eta(i)}^D(t)\right|\\
 \end{align*}
 {
Because $\eta$ is arbitrary, we take the infimum over $\eta$ to retrieve
\[\|P(C,\psi,\Sigma)-P(D,\psi,\Sigma)\|_1\le\kappa_{\infty}(\psi, C, D) W_{1}(C,D)+ (L_{\infty}^C \land L_{\infty}^D) \delta_{1}(\psi,C,D).\]}
 This completes the proof of \eqref{equ:main estimate W1 dist}.
\end{proof}

\Cref{thm:main stability thm} provides a general bound on the difference of two PCs with $T=\Sigma$ with respect to the bottleneck and 1-Wasserstein distances. However, due to the general nature of the PC framework, \Cref{thm:main stability thm} does not offer insight into the implied stability of specific curves. Rather, it offers a simple way to perform a stability analysis on any specific curve.  In the next section, we offer examples by applying \Cref{thm:main stability thm} to lifespan-based curves.

\section{Explicit Bounds for Lifespan-based Curves}

\label{sec:application of main thm}

The main focus of this section is twofold. First, we apply \Cref{thm:main stability thm} to investigate the stability of given PCs. We will take lifespan curves as illustration for this discussion.  Second, we introduce two families of PCs that we call {\bf normalized} and {\bf entropy-based} persistence curves.  In consideration of readability, we postpone the stability discussions for the other curves listed in \Cref{tab:PC} to \Cref{app:stability}. 
 
To apply \Cref{thm:main stability thm}, we must control  $\kappa_1$, $\kappa_{\infty}$, $\delta_1$, $\delta_{\infty}$ (defined in \Cref{tab:notations}).  The value of $\kappa_1$ and $\kappa_\infty$ relate to maximum function values of the user-defined $\psi$.  It is the bounds on $\delta_1$ and $\delta_\infty$ that might lead to the stability result. 
We investigate stability of PCs with respect to both $W_1$ and $W_{\infty}$ distances.  In some cases, additional assumptions are required to establish the stability, while in other cases, stability holds without further assumptions. To highlight their differences, we call a PC conditionally stable if additional assumptions are required.  In practice, it is also reasonable to consider a subspace of $\mathcal{D}$ whose smallest birth value and lifespan are uniformly bounded below, and the largest death value is uniformly bounded above, i.e. consider $\mathcal{D}_{M,m,q} = \{ D\in\mathcal{D}~|~m\leq b<d\leq M,~d-b\ge q >0\}$. Digital image analysis with sublevel set filtration is one such situation this restriction arises naturally with $m = 0, M = 255$, and $q=1$. 
\subsection{Basic Curves}
\label{subsec:lifespan}
An intuitive way to create PCs is to select $\psi(b,d)$ based on usual diagram point statistics such as lifespan or midlife.  In what follows, we demonstrate stability analysis of the lifespan curve defined in \Cref{ex:lifespan curve} as $\mathbf{l}(D)\equiv P(D,d-b,\Sigma)$ using \Cref{thm:main stability thm}. 

The computations for $\kappa_1 = L^C+L^D\le 2(L^C\lor L^D)$ and $\kappa_\infty = L_\infty^C+L_\infty^D\le 2(L_\infty^C\lor L_\infty^D) $ are straightforward from the definitions of \Cref{tab:notations} and do not rely on the {matching.  Now, let $\eta$ be a matching. Similar to the proof of Theorem \ref{thm:main stability thm}, we will order the points accordingly. Then we obtain}  
\begin{align*}
     \max_{1\le i\le n_{\eta}}|\psi^C(b_i,d_i)-\psi^D(\eta_{b_i},\eta_{d_i})|& =\max_{1\le i\le n_{\eta}}|(d_i-b_i)-(\eta_{d_i}-\eta_{b_i})|\\ & \le\max_{1\le i\le n_{\eta}}|(d_i-\eta_{d_i})|+|(b_i-\eta_{b_i})|
\end{align*}
{Take infimum over $\eta$ to obtain $\delta_{\infty}(\psi, C, D) \leq 2W_{\infty}(C,D)$.  For $\delta_{1}(\psi, C, D)$, we consider}
\begin{align*}
     \sum_{i=1}^{n_{\eta}}|\psi^C(b_i,d_i)-\psi^D(\eta_{b_i},\eta_{d_i})|
    &=\sum_{i=1}^{n_{\eta}}|(d_i-b_i)-(\eta_{d_i}-\eta_{b_i})|\\
    &\le\sum_{i=1}^{n_{\eta}} |(d_i-\eta_{d_i})|+|(b_i-\eta_{b_i})|
\end{align*}
{Take infimum over $\eta$ to obtain $\delta_1(\psi, C, D) \leq 2W_1(C,D)$.}
Therefore by \Cref{thm:main stability thm} we conclude
\begin{align}
    \|\mathbf{l}(C)-\mathbf{l}(D)\|_1&\le 2(L^C\lor L^D)W_\infty(C,D) + 2(L^C\land L^D)W_\infty(C,D) = 2(L^C+L^D)W_\infty(C,D), \label{equ:lifespan Winfty bound}\\
    \|\mathbf{l}(C)-\mathbf{l}(D)\|_1&\le 2(L_\infty^C\lor L_\infty^D)W_1(C,D) + 2(L_\infty^C\land L_\infty^D)W_1(C,D) = 2(L_\infty^C+L_\infty^D)W_1(C,D).\label{equ:lifespan W1 bound}
\end{align}
For \eqref{equ:lifespan Winfty bound}, the bound can become arbitrarily large without restricting the lifespan and number of points of a persistence diagram.  $\mathbf{l}$ can be stable with respect to $W_{\infty}$ only by controlling the total lifespan in a diagram.  This is, however, not the case that occurs naturally leading us to conclude that $\mathbf{l}$ is not stable with respect to $W_\infty$.

On the other hand, the bound \eqref{equ:lifespan W1 bound} is much more manageable and most importantly depends only on the largest lifespan, which is commonly globally bounded in applications. In other words, if we further assume that $C,~D\in\mathcal{D}_{M,m,q}$, then \eqref{equ:lifespan W1 bound} becomes
$ \|\mathbf{l}(C)-\mathbf{l}(D)\|_1 \leq 4(M-m)W_1(C,D)$.  Thus, $\mathbf{l}$ is conditionally stable with respect to $W_1$.

{
}

\subsection{Normalized persistence curve}
\label{subsec:normalizedlifespan} 
Motivated by observations in \Cref{subsec:lifespan}, in order to obtain a stable summary, we must have a way to control $\delta$. To this end, we propose here a simple modification of basic PCs.  Despite its simplicity, the following modification possesses both theoretical and practical advantages. 
\begin{definition}
\label{def:normalized curve}
{Given $D\in\mathcal{D}$, let $\phi(D,b,d,t)=\phi(D;b,d)$ be a function that does not depend on $t$.  Assume that $D\neq \Delta$ and $\phi$ is not a zero function. Define \[\psi(D;b,d) = \frac{\phi(D;b,d)}{\sum_{(b',d')\in D}|\phi(D;b',d')|}.\] 
Then the \textbf{normalized persistence curve} with respect to $\phi$ is defined 
to be $P\left(D,\psi,\Sigma\right)$. In the case that $D = \Delta$, we definte the normalized curve to be the 0 function.}
\end{definition}

An immediate observation is that the absolute value of function values of a normalized PC lie between $0$ and $1$. More importantly, because of the normalization factor, one can show that for the normalized PCs, $\kappa$ is uniformly bounded by 2, i.e.
\begin{equation}
\label{equ:sl kappas}
   \kappa_1(\psi, C, D) \leq  2{\text{, and }} \kappa_{\infty}(\psi, C, D)\leq 2.
\end{equation}
Thus, to investigate the stability of the normalized PCs, it remains to control $\delta$.

As an illustration, we now consider the lifespan curves. 
According to the \Cref{def:normalized curve}, the \textbf{normalized lifespan curve} is \[\mathbf{sl}(D) \equiv P\left(D,\frac{d-b}{L^D},\Sigma\right).\]
We will show that the normalized lifespan curve is not only more stable than the lifespan curve, but also offers better performance in applications.

{We provide the detailed bounds on $\delta$ for $\mathbf{sl}$. Let $\eta$ be any matching between $C$ and $D$.  To ease notation we will let $\ell_{i} = {d_i}-{b_i}$ and $\ell_{\eta_i} = \eta_{d_i}-\eta_{b_i}$.  Assume, without loss of generality that $L^C\le L^D$, i.e. $L^C \lor L^D = L^D$.  We also note that since $|L^C-L^D|\le\sum_{i=1}^{n_\eta}|d_i-\eta_{d_i}|+|b_i-\eta_{b_i}|\le 2(n^C+n^D) \max_{i} (|d_i-\eta_{d_i}|\lor|b_i-\eta_{b_i}| )$, by taking infimum over $\eta$, one has $|L^C-L^D|\leq 2(n^C+n^D) W_{\infty}(C,D)$. Then
\begin{align*}
    \max_{1\le i\le n_{\eta}}|\psi^C(b_i,d_i)-\psi^D(\eta_{b_i},\eta_{d_i})|
   & =\max_{1\le i\le n_{\eta}}\left|\frac{\ell_i}{L^C}-\frac{\ell_{\eta_i}}{L^D}\right|\\
    &\le\max_{1\le i\le n_{\eta}}\frac{|\ell_i-\ell_{\eta_i}|}{L^D} +\max_{1\le i\le n_{\eta}}\ell_i\frac{|L^C-L^D|}{L^CL^D}\\
    &\le \frac{1}{L^C\lor L^D}\max_{1\le i\le n_{\eta}}{|\ell_i-\ell_{\eta_i}|} + \frac{(L^C_\infty\lor L^D_\infty)|L^C-L^D|}{L^C L^D} \\
    &\leq \frac{1}{L^C\lor L^D}\max_{1\le i\le n_{\eta}}{|\ell_i-\ell_{\eta_i}|} + \frac{(L^C_\infty\lor L^D_\infty)2(n^C+n^D)W_{\infty}(C,D)}{L^C L^D}.
\end{align*}
Therefore, by taking infimum over $\eta$, we have
\begin{equation}
\label{equ:sl deltainf bound}
\delta_\infty(\psi, C, D) \leq  \frac{2W_\infty(C,D)}{L^C\lor L^D} +\frac{(L^C_\infty\lor L^D_\infty)2(n^C+n^D)W_{\infty}(C,D)}{L^C L^D}. 
\end{equation}

Similarly, for a matching $\eta:C\to D$, we have $|L^C-L^D|\le\sum_{i=1}^{n_{\eta}}|d_i-\eta_{d_i}|+|b_i-\eta_{b_i}|$ and take infimum over $\eta$ to obtain $|L^C-L^D|\leq 2W_1(C,D)$. Thus,
\begin{align*}
    \sum_{i=1}^{n_{\eta}}\left|\psi^C(b_i,d_i)-\psi^D(\eta_{b_i},\eta_{d_i})\right|
    &=\sum_{i=1}^{n_{\eta}}\left|\frac{\ell_i}{L^C}-\frac{\ell_{\eta_i}}{L^D}\right|\\
    &\le\sum_{i=1}^{n_{\eta}}\frac{|\ell_i-\ell_{\eta_i}|}{L^D} +\sum_{i=1}^{n_{\eta}}\frac{\ell_i}{L^C}\frac{|L^C-L^D|}{L^D}
    \\&\le \sum_{i=1}^{n_{\eta}}\frac{|b_i-\eta_{b_i}|+|d_i-\eta_{d_i}|}{L^C\lor L^D } + \frac{|L^C-L^D|}{L^C\lor L^D}\\
    & \leq \frac{1}{L^C\lor L^D} \left[\sum_{i=1}^{n_{\eta}}{|b_i-\eta_{b_i}|+|d_i-\eta_{d_i}|}\right] + \frac{2W_1(C,D)}{L^C\lor L^D}.
\end{align*}
By taking infimum over $\eta$, we have
\begin{equation}
\label{equ:sl delta1 bound}
\delta_1(\psi, C, D) \leq  \frac{4W_1(C,D)}{L^C\lor L^D}.
\end{equation}

We note that $L^C L^D = (L^C \land L^D ) (L^C \lor L^D)$.  Also, $\frac{L^C_{\infty} \land L^D_{\infty}}{L^C\lor L^D} \leq \frac{L^C_{\infty}}{L^C} \land \frac{L^D_{\infty}}{L^D} $ by the following elementary calculation, \Cref{lem:simply number lem}.
\begin{lemma}
\label{lem:simply number lem}
Let $a,b,c,d$ be positive real numbers.  Then the following inequality holds. 
\begin{equation}
    \frac{a\land b}{c \lor d} \leq \frac{a}{c} \land \frac{b}{d}.
\end{equation}
\end{lemma}
\begin{proof}
Without loss of generality, suppose $a \land b = a$. We show (1) $\frac{a}{c \lor d} \leq \frac{a}{c}$ and (2) $\frac{a}{c \lor d} \leq \frac{b}{d}.$ We get (1) immediately since $c\lor d \ge c$. For (2), notice $\frac{a}{c \lor d}\le \frac{b}{c \lor d}$. We conclude (2) by noting that $c \lor d \ge d$. Therefore $\frac{a\land b}{c \lor d} \leq \frac{a}{c} \land \frac{b}{d}.$ 
\end{proof}

}
Therefore by \eqref{equ:sl kappas}, \eqref{equ:sl deltainf bound},  \eqref{equ:sl delta1 bound}, and \Cref{thm:main stability thm} we conclude
\begin{align}
    \|\mathbf{sl}(C)-\mathbf{sl}(D)\|_1&\le 2W_\infty(C,D) + (L^C\land L^D)\left(\frac{2W_\infty(C,D)}{L^C\lor L^D} + \frac{2(n^C+n^D)(L^C_\infty\lor L^D_\infty)W_\infty(C,D)}{L^C L^D}\right)\nonumber\\
    &=2W_\infty(C,D)\left(2+(n^C+n^D)\frac{L_\infty^C\lor L_\infty^D}{L^C\lor L^D}\right) \label{equ:normalized life Winfty bound}; \\
    \|\mathbf{sl}(C)-\mathbf{sl}(D)\|_1&\le 2W_1(C,D) + L_\infty^C\land L_\infty^D\frac{4W_1(C,D)}{L^C\lor L^D} \leq 2W_1(C,D)+ \frac{L^C_{\infty}}{L^C} \land \frac{L^D_{\infty}}{L^D} 4W_1(C,D) \nonumber\\& \le 6W_1(C,D).\label{equ:normalized life W1 bound}
\end{align}
 The bound \eqref{equ:normalized life Winfty bound} is better than \eqref{equ:lifespan Winfty bound} because both denominator and numerator have a factor of $n$ in \eqref{equ:normalized life Winfty bound}.  If we further assume that $C,~D \in \mathcal{D}_{M,m,q}$, then \eqref{equ:normalized life Winfty bound} becomes
\begin{equation}
     \| \mathbf{sl}(C) - \mathbf{sl}(D) \|_1 \le\left[2\frac{M-m}{q} + 4 \right] W_{\infty}(C,D),~\forall C,~D\in\mathcal{D}_{M,m,q}.
\end{equation}
Thus, we conclude that $\mathbf{sl}$ is conditionally stable in $W_{\infty}$.  \eqref{equ:normalized life W1 bound} implies that $\mathbf{sl}$ is stable in $W_1$.

When comparing stability result of $\mathbf{l}$ and $\mathbf{sl}$, we observe that adding the normalization factor ``stabilizes'' the lifespan curve: $\mathbf{l}$ is not stable with respect to $W_{\infty}$, while $\mathbf{sl}$ is conditionally stable with respect to $W_{\infty}$;  $\mathbf{l}$ is conditionally stable with respect to $W_1$, while $\mathbf{sl}$ is stable with respect to $W_1$.  A natural question is whether adding the normalization factor would always stabilize a given persistence curve.  The answer is positive with a mild condition.  We devote the rest of this section to this discussion.  To formulate this problem, we consider the case where
\begin{itemize}
    \item[(A1)]$\phi(b,d,t)=\phi(b,d)$ is independent of $t$;
    \item[(A2)]$\phi$ satisfies $\delta_1(\phi, C, D) \leq K_1 W_1(C,D)$ where $K_1$ is a uniform constant.
\end{itemize}
We will focus on the $W_1$ stability.  By (A1) and (A2) and from Table~\ref{tab:notations}, we find that $\kappa_{\infty}(\phi, C, D) = \max_{(b,d)\in C}\phi(b,d) + \max_{(b,d)\in D} \phi(b,d)$. 
By Theorem~\ref{thm:main stability thm}, the Lipschitz bound for the generic (without normalizing) persistence curve $ P(C, \phi, \Sigma)$ is
\begin{equation}
\label{equ:pc w1 bound}
    \| P(C, \phi, \Sigma) - P(D, \phi, \Sigma) \|_1 \leq \left[\max_{(b,d)\in C}\phi(b,d) + \max_{(b,d)\in D} \phi(b,d) + K_1(L_{\infty}^C \land L_{\infty}^D) \right] W_1(C,D).
\end{equation}
We observe that this class of persistence curves is at best conditionally stable because one has to assume a uniform bound for $\max_{(b,d)\in C}\phi(b,d)$ and $L_{\infty}^C \land L_{\infty}^D$.

On the other hand, we now apply the normalization factor and consider its persistence curve $P\left(D, \frac{\phi}{\Phi} , \Sigma \right)$, where $\Phi(D) := \sum_{(b,d)\in D} |\phi(b,d)|$. The Lipschitz bound for $P\left(D, \frac{\phi}{\Phi} , \Sigma \right)$ is summarized below whose proof can be found at the end of this subsection.
\begin{theorem}
\label{thm:normalized pc stability}
Let $\phi$ satisfy (A1) and (A2).  Then the normalized persistence curve $P(\cdot,\frac{\phi}{\Phi} , \Sigma )$ satisfies
\begin{equation}
\label{equ:normalized pc w1 bound}
\left\|P\left(C, \frac{\phi}{\Phi} , \Sigma \right) - P\left(D, \frac{\phi}{\Phi}, \Sigma\right)\right\|_1\le
\left[2 + 2K_1 \frac{L_{\infty}^C }{\Phi(C)} \lor \frac{L_{\infty}^D }{\Phi(D)} \right] W_{1}(C,D).    
\end{equation}
\end{theorem}
We can now compare the bounds \eqref{equ:pc w1 bound} and \eqref{equ:normalized pc w1 bound}.  As stated above, \eqref{equ:pc w1 bound} is at best conditional stable; on the other hand, \eqref{equ:normalized pc w1 bound} can be stable.  In fact, the ratio $ \frac{L_{\infty}^C }{\Phi(C)} \lor \frac{L_{\infty}^D }{\Phi(D)} $ can be bounded uniformly by a constant for a wide class of functions. Intuitively speaking, this ratio can be thought as ``$\infty$-norm over 1-norm'' since the numerator is the largest lifespan while the denominator is the total sum of the absolute values of $\phi$.  For instance, consider $\phi(b,d)=d-b$ as discussed in \eqref{equ:normalized life W1 bound} and hence, $\Phi(C) = L^C$.  Therefore, $\frac{L^C_\infty}{\Phi(C)} = \frac{L^C_\infty}{L^C} \leq 1 $. Another example could be $\phi(b,d) = \tan^{-1}(d-b)$ used in \cite{kusano2016persistence}.  

To conclude, we have shown that for a class of functions $\phi$, normalization stabilizes a given persistence curve.  To specify the class of functions, consider
\begin{itemize}
    \item[(A3)] The function $\phi$ satisfies the property that there exists some positive real number $K_2$ so that $\frac{L^C_\infty}{\Phi(C)}\le K_2$ for any $C\in \mathcal{D}$.
\end{itemize}
Let $\phi$ satisfy (A1), (A2), and (A3). The normalization factor would always stabilize the persistence curves.  More specifically,
\begin{equation}
    P\left(C, \frac{\phi}{\Phi} , \Sigma \right) \text{ is stable but } P(C, \phi, \Sigma) \text{ is conditionally stable}.
\end{equation}




Lastly, we present the proof of Theorem~\ref{thm:normalized pc stability} below.
\begin{proof}[Proof of Theorem~\ref{thm:normalized pc stability}:]
Now, with $C,D\in\mathcal{D}$,and $\phi(D,b,d)$ given, define $\Phi(D) = \sum_{(b,d)\in D} |\phi(b,d)|$. Let $\eta:C\to D$ be a matching. Denote $\phi(C,b_i,d_i)$ as $\phi_i$ and $\phi(D,\eta_{b_i},\eta_{d_i})$ as $\phi_{\eta_i}$. Consider
\begin{align*}
    \sum_{i=1}^{n_{\eta}}  \left|\frac{\phi_i}{\Phi(C)} - \frac{\phi_{\eta_i}}{\Phi(D)}\right|&  \leq  \sum_{i=1}^{n_{\eta}} \frac{|\phi_i - \phi_{\eta_i}|}{\Phi(D)} + \sum_{i=1}^{n_{\eta}}|\phi_i| \frac{|\Phi(C) - \Phi(D)|}{\Phi(C) \Phi(D)}\\
    &=  \frac{1}{\Phi(D)}\sum_{i=1}^{n_{\eta}}|\phi_i - \phi_{\eta_i}| + \frac{1}{\Phi(D)} |\Phi(C) - \Phi(D)|.
\end{align*}
By the same argument, one may also obtain
\begin{equation*}
     \sum_{i=1}^{n_{\eta}}  \left|\frac{\phi_i}{\Phi(C)} - \frac{\phi_{\eta_i}}{\Phi(D)}\right| \leq  \frac{1}{\Phi(C)}\sum_{i=1}^{n_{\eta}}|\phi_i - \phi_{\eta_i}| + \frac{1}{\Phi(C)} |\Phi(C) - \Phi(D)|.
\end{equation*}
Therefore, we have
\begin{align*}
    \sum_{i=1}^{n_{\eta}}  \left|\frac{\phi_i}{\Phi(C)} - \frac{\phi_{\eta_i}}{\Phi(D)}\right| &\leq \frac{1}{\Phi(C) \lor \Phi(D)} \left[\sum_{i=1}^{n_{\eta}}|\phi_i - \phi_{\eta_i}|+  |\Phi(C) - \Phi(D)|\right]\\
    &\leq \frac{2}{\Phi(C) \lor \Phi(D)} \left[\sum_{i=1}^{n_{\eta}}|\phi_i - \phi_{\eta_i}|\right].
\end{align*}
Note that $|\Phi(C) - \Phi(D)|\leq \sum_{i=1}^{n_{\eta}}|\phi_i - \phi_{\eta_i}|$ by the reverse triangle inequality.

By taking infimum over $\eta$ and then by (A2), we have
\begin{equation*}
    \delta_1(\frac{\phi}{\Phi}, C,D) \leq \frac{2}{\Phi(C) \lor \Phi(D)} \delta_1(\phi, C, D) \leq \frac{2K_1 W_1(C,D)}{\Phi(C) \lor \Phi(D)}.
\end{equation*}
Thus, by Theorem~\ref{thm:main stability thm}, Lemma~\ref{lem:simply number lem}, and $\kappa_{\infty}\leq 2$
\begin{align*}
    \left\|P\left(C, \frac{\phi}{\Phi}\right) - P\left(D, \frac{\phi}{\Phi}\right)\right\|_1 &\leq 2 W_1(C,D) + \frac{L_{\infty}^C \land L_{\infty}^D}{\Phi(C) \lor \Phi(D)} 2K_1W_1(C,D) \\
    &= \left[2 + \frac{(2K_1)(L_{\infty}^C \land L_{\infty}^D)}{\Phi(C) \lor \Phi(D)}\right] W_{1}(C,D) \\
    &\leq \left[2 + 2K_1 \frac{L_{\infty}^C }{\Phi(C)} \lor \frac{L_{\infty}^D }{\Phi(D)} \right] W_{1}(C,D).
\end{align*}
This completes the proof.
\end{proof}

\subsection{Entropy-based persistence curve}
 Motivated by the life entropy curve discussed in \Cref{ex:persistent entropy curve},
we propose another modification of basic PCs based on the idea of the entropy.   In view of the PC framework, we could generalize the format of the life entropy curve. 
\begin{definition}
\label{def:entropy-based persistence curve}
Given $D\in \mathcal{D}$, let $\phi(D,b,d,t)=\phi(D,b,d)$ be a function that does not depend on $t$.
Then the \textbf{entropy-based persistence curve} with respect to $\phi$ is defined as \[P\left(D,-\frac{\phi(D,b,d)}{\sum_{(b',d')\in D}\left|\phi(D,b',d')\right|}\log \frac{\phi(D,b,d)}{\sum_{(b',d')\in D}\left|\phi(D,b',d')\right|},\Sigma\right).\]
\end{definition}
An immediate computation leads to $\kappa_1 \le\log n^C+ \log n^D \le 2\log (n^C\lor n^D)$ and $\kappa_\infty\le \frac{2}{e}$. Similar to discussion in \Cref{subsec:normalizedlifespan}, to investigate the stability, it remains to obtain bounds on $\delta$.

As an illustration, we consider the lifespan entropy curve discussed in \Cref{ex:persistent entropy curve}, $\mathbf{le}(D) := P(D, -\frac{d-b}{L^D}\log\frac{d-b}{L^D}, \Sigma)$. To establish the stability result for $\mathbf{le}$, additional assumption is required because of the following result. Before the computations, we note that $\lim_{x\rightarrow 0^+} x\log x = 0$, so we may define $0\log 0$ to be $0$.  Moreover, $\max_{x\in [0,1]}(-x\log x )= \frac{1}{e}$ and $\min_{x\in [0,1]} (-x\log x) = 0$.
\begin{lemma}\label{lem:loglemma}
Let $0\le x,y\le 1$
If $|x-y|\le\epsilon\le \frac{1}{e}$ then $|x\log x-y\log y|\le -\epsilon\log{\epsilon}.$
\end{lemma}
\begin{proof}
$h(x,y) =-|x-y|\log|x-y|-|x\log x-y\log y|$. Notice that $h$ is symmetric about the line $y=x$. Thus without loss of generality, we may assume $x\ge y$. Fix some $y\in [0,1]$. We obtain a function of $x$, $f^y(x) = h(x,y)$. We will show that for every $y\in[0,1]$, $f^y(x)\ge 0$ whenever $x-y\le\frac{1}{e}$. This will then allow us to conclude by the symmetry of $h$ that $h(x,y)\ge 0$ for all $x,y\in[0,1]$ with $|x-y|\le \epsilon\le \frac{1}{e}$. We have three cases to consider: a) $y\in[0,1-\frac{1}{e})$; b) $y\in[\frac{1}{e},1)$; c) $y=1$.

a) If $y\in[0,1-\frac{1}{e})$ then $f^y$ is defined on the domain $[y,y+\frac{1}{e}]$. Notice that $f^y(y) = 0$ and $f^y(y+\frac{1}{e}) = \frac{1}{e}\log\frac{1}{e} - |x\log x - y\log y| \ge 0$ since $\frac{1}{e}\log\frac{1}{e}$ is the maximum of $|x\log x - y\log y|$ on the interval from $[0,1]$. Finally, the derivative of $f^y$, reveals that $f^y$ takes a maximum on the interval $(y,y+\frac{1}{e})$, thus $f^y(x)\ge 0$ for $y\in[0,1-\frac{1}{e})$.

b)If $y\in[\frac{1}{e}, 1)$, $f^y$ is defined on the interval $[y,1]$. Again we see $f^y(y) = 0$ and also $f^y(1) = 0$. The derivative reveals a maximum on $(y,1)$ and thus $f^y\ge 0$ for all $y\in[1-\frac{1}{e},1)$

c) Finally, if $y=1$ then $f^y(1) = 0$.

Therefore, we've shown that $h(x,y)\ge 0$ for $0\le y\le x \le 1$ and $|x-y|\le \frac{1}{e}$. Since $h$ is symmetric, we see $h(x,y)\ge 0$ for $0\le y, x \le 1$ and $|x-y|\le\epsilon\le \frac{1}{e}$ as desired.
\end{proof}

In connection to the stability analysis, we follow \cite{persistentEntropyStability} by defining the relative error between two persistence diagrams as 
\begin{equation*}
    r_p(C,D) = \frac{2(n^C\lor n^D)^{1-\frac{1}{p}}W_p(C,D)}{L^C\lor L^D} {\text{, and } r_\infty(C,D) = \frac{2(n^C\lor n^D)W_\infty(C,D)}{L^C\lor L^D}}.
\end{equation*}
Let $C,D$ be diagrams and $\eta:C\to D$ be a matching. Order the points in a similar way to the proof of Theorem \ref{thm:main stability thm}. For $(b_i,d_i)\in C$ we denote the quantity $d_i-b_i$ by $\ell_i$. Similarly for $(\eta_{b_i},\eta_{d_i})\in D$, we denote $\eta_{d_i}-\eta_{b_i}$ by $\ell_{\eta_i}$.
\begin{align}
    \max_{1\le i\le {n_{\eta}}}\left|\frac{\ell^C_i}{L^C}-\frac{\ell_{\eta_i}}{L^D}\right|&\le\max_{1\le i\le {n_{\eta}}}\frac{|\ell_i-\ell_{\eta_i}|}{L^D} +\max_{1\le i\le {n_{\eta}}}\ell_i\frac{|L^C-L^D|}{L^CL^D} \nonumber \\
    &\le \frac{2W_\infty(C,D)}{L^C\lor L^D} + \frac{2{n_{\eta}}W_\infty(C,D)}{(L^C\lor L^D)} \le 2r_{\infty}(C,D). \label{equ:delta step est}
\end{align}
Thus, if $r_\infty(C,D)\le \frac{1}{2e}$, by \eqref{equ:delta step est} and \Cref{lem:loglemma}, we obtain
\[\delta_\infty(\psi, C, D)\le\max_{1\leq i \leq {n_{\eta}}} \left|\frac{l_i^C}{L^C}\log(\frac{l_i^C}{L^C}) - \frac{l_{\eta_i}^D}{L^D}\log(\frac{l_{\eta_i}^D}{L^D})\right| \le -2r_\infty(C,D)\log{2r_\infty(C,D)}. \]
If the indexing is instead for $W_1(C,D)$, then
\begin{align*}
    \sum_{i=1}^{n_{\eta}}\left|\frac{\ell^C_i}{L^C}-\frac{\ell_{\eta_i}^D}{L^D}\right|&\le \frac{4W_1(C,D)}{L^C\lor L^D} = 2r_1(C,D).
\end{align*}
If $r_1(C,D)\le\frac{1}{2e}$, \Cref{lem:loglemma} guarantees
\begin{align*}
    \delta_1(\psi, C, D) &\le \sum_{i=1}^{n_{\eta}} \left|\frac{l_i^C}{L^C}\log(\frac{l_i^C}{L^C}) - \frac{l_{\eta_i}^D}{L^D}\log(\frac{l_{\eta_i}^D}{L^D})\right|  \\
    &\le-\sum_{i=1}^{n_{\eta}} 2r_1(C,D)\log{2r_1(C,D)}\le -2(n^C\lor n^D)r_1(C,D)\log{2r_1(C,D)}.
\end{align*}
Therefore, if $r_\infty(C,D)\le \frac{1}{2e}$ and $r_1(C,D)\le \frac{1}{2e}$ respectively, then by \Cref{thm:main stability thm}
\begin{align}
    \nonumber\|\mathbf{le}(C)-\mathbf{le}(D)\|_1&\le 2\log (n^C\lor n^D) W_\infty(C,D)\\&- (L^C\land L^D)\frac{4(n^C\lor n^D)W_\infty(C,D)}{L^C\lor L^D}\log{\frac{4(n^C\lor n^D)W_\infty(C,D)}{L^C\lor L^D}}\label{equ:lifespan entropy Winfty bound}, \\
    \|\mathbf{le}(C)-\mathbf{le}(D)\|_1&\le  \left(\frac{2}{e} - (L_\infty^C\land L_\infty^D)2(n^C\lor n^D)\frac{2}{L^C\lor L^D}\log{\frac{4W_1(C,D)}{L^C\lor L^D}}\right) W_1(C,D). \label{equ:lifespan entropy W1 bound}
\end{align}
We see both of these bounds require tighter control over both the number of points in the diagram and the allowable lifespans. For example, the requirement that $r_\infty(C,D)\le \frac{1}{2e}$ implies that $4eW_\infty(C,D)\le \frac{L^C\lor L^D}{(n^C\lor n^D)}$. We can think of this as being a requirement that a multiple of the bottleneck distance is bounded above by the average lifespan (though the values in the numerator and denominator may correspond to different diagrams). Thus, the life entropy curve is conditionally stable with respect to both $W_\infty$ and $W_1$. \Cref{equ:lifespan entropy Winfty bound} is similar to the bound appearing in \cite{persistentEntropyStability}. The differences are that we make use of the natural log whereas they make use of a base-2 log; \eqref{equ:lifespan entropy W1 bound} is the new control that we provide.

\subsection{Bounds on Persistence Landscapes}
\label{subsec:bounds on PL}
At this point, we have discussed the stability analysis of PCs with $T=\Sigma$.  { One natural question is that can one extend the stability analysis to the case when $T\ne \Sigma$?  It is, in general, a more difficult task. For instance, if $T = \max$, then the corresponding persistence curves might be unstable as shown below.
\begin{example}
Let $\psi(b,d) = d-b$. For any $n\in\mathbb{N}$ and $n>1$, define $C_n = \{(0,n)\}$ and $D_n = \{(0,n-1)\}$. Clearly $W_1(C_n,D_n) = 1$. However, 
\begin{align*}
    \|P(C_n,\psi,\max) - P(D_n,\psi,\max)\|_1 & = \| n \chi_{[0,n)} - (n-1) \chi_{[0, n-1)} \|_1 \\
                                              & = \int_{0}^{n-1}n-(n-1)\mathrm{d}t + \int_{n-1}^n n\mathrm{d}t = 2n-1.
\end{align*}
Thus, we find that for $n > 1$,
\begin{equation}
     \|P(C_n,\psi,\max) - P(D_n,\psi,\max)\|_1 > W_1(C_n,D_n).
\end{equation}
Hence, the persistence curve using the lifespan function and max statistic is not stable.
\end{example}
On the other hand, it is well-know that the persistence landscapes which are $P(D, \psi, \max)$ are stable.
}
In this subsection, we take persistence landscapes as an illustration to discuss stability analysis for those PCs with $T\ne \Sigma$.

It is important to note that the \Cref{thm:main stability thm} does not apply in this case.  However, using this framework, we provide an elementary proof of the stability result (Theorem 12 in \cite{bubenik2015statistical}) of persistence landscapes which states
\begin{equation}
    \label{equ:PL stability}
    \| P(C, \psi, \max ) - P(D, \psi, \max ) \|_{\infty} \leq W_{\infty}(C,D).
\end{equation}
The main idea of this elementary proof is to establish a similar result to \Cref{lem:estimate}. Specifically, we follow the similar argument to \eqref{equ:diff indi func real} given two continuous functions $\psi_1$, $\psi_2$, and two finite intervals $I$, $J\subset\mathbb{R}$, our main focus is to estimate $\|\psi_1 \chi_{I}-\psi_2\chi_J\|_{\infty}$
\begin{align}
&\max\{|\psi_1(t) \chi_{I}(t)-\psi_2(t)\chi_J(t)|\}\nonumber \\
&= \bigg(\max_{t\in I\setminus J} |\psi_1(t)|\bigg) \lor \bigg( \max_{t\in I\cap J} | \psi_1(t) - \psi_2(t)| \bigg) \lor \bigg( \max_{t\in J\setminus I} | \psi_2(t)|\bigg). \label{equ:diff max indi func real}
\end{align}

\begin{lemma}
\label{lem:lemma for pl}
Let $f(t) =  | \psi_1(t) \chi_{[b_1,d_1)}(t) -\psi_2(t) \chi_{[b_2,d_2)}(t) |$, where $\psi_i(t) = \min\{ t-b_i, d_i - t \}$ for $i = 1,~2$.  Then
\begin{equation}
\label{equ:pl est}
    \|f\|_{\infty} \leq |d_2-d_1| \lor |b_2 - b_1|.
\end{equation}
\end{lemma}
\begin{proof}
We outline the idea of the proof.  Similar to the proof of \Cref{lem:estimate}, there are three cases to consider.  In each case, it remains to prove that each term in \eqref{equ:diff max indi func real} is bounded above by $|d_2-d_1| \lor |b_2 - b_1|$.  To see that, we will use the fact that $\max_{t\in [b,d)} \min\{{t-b, d-t}\} = \frac{d-b}{2}$ when $t = \frac{b+d}{2}$ and exhaust all possible locations of $\frac{b+d}{2}$.  The details of the proof are included in the \Cref{app:landscapes}.
\end{proof}
Let $\eta$ be a matching. Thus, to prove \eqref{equ:PL stability}, by \Cref{lem:lemma for pl} we obtain
\begin{align*}
    &\| P(C, \psi, \max) - P(D, \psi, \max) \|_{\infty} \\
    &= \max_{1\leq i \leq {n_{\eta}}} \bigg[ \max_{t \in [b_i, d_i) \cup [\eta_{b_i}, \eta_{d_i})} | \psi(b_i, d_i, t) \chi_{[b_i,d_i)}(t) -\psi(\eta_{b_i}, \eta_{d_i}, t) \chi_{[\eta_{b_i},\eta_{d_i})}(t) | \bigg] \\
    &\leq \max_{1\leq i \leq {n_{\eta}}} |d_i-\eta_{d_i}| \lor |b_i - \eta_{b_i}|.
\end{align*}
By taking infimum over $\eta$, we obtain
\begin{equation*}
    \| P(C, \psi, \max) - P(D, \psi, \max) \|_{\infty} \leq W_{\infty}(C,D).
\end{equation*}

Due to the generality of the PC framework, stability (Quality 2) depends on the choice of $\psi$ and $T$. We have provided several examples of stable, conditionally stable, and unstable curves in this section. In the next section, we show the computational efficiency, efficacy, and experimental stability of our proposed curves.

\section{Computation and Applications}
\label{sec:computation and applications}
In this section, we explore computational aspects of PCs.
First, we demonstrate the implementation of a general PC, and discuss its efficiency.  Second, we apply PCs to two applications: parameter determination for a discrete dynamical system and image texture classification.  Comparisons among results by PCs and other TDA methods are also included.  Third, we explore how PCs handle noise in practice.  Finally, we discuss some limitations of PCs and persistent homology in general.

\subsection{Implementation and Efficiency}
	\renewcommand{\algorithmicrequire}{\textbf{Input:}}
	\renewcommand{\algorithmicensure}{\textbf{Output:}}
	\begin{algorithm} \caption{Pseudo-code for computing PCs. \label{alg1}}
		\begin{algorithmic}[1] 
			\Require Persistence Diagram $D$, initial value $m$, terminated value $M$, user-defined function $\psi(D;b,d,t)$, statistic $T$, number of mesh points $N$.
			\Ensure A vector $V\in\mathbb{R}^{n}$ values of $P(D,\psi,T)(\{m+k\frac{M - m}{N}\}_{k=1}^{N})$
			\For{$1\leq k\leq N$}
			\State $t_k\leftarrow m+k\frac{M-m}{N}$.  \Comment{Compute the mesh points.}
			\State $D_{t_k}\leftarrow D\cap F_{t_k}$. \Comment{ Find the fundamental box.}
			\State $V[k] \leftarrow T(\psi(D_{t_k}))$. \Comment{Evaluate $\psi$ at the fundamental box and summarize the result by $T$.}
			\EndFor
			\State\Return V
		\end{algorithmic}
	\end{algorithm}

	\begin{algorithm} \caption{Vectorized algorithm for computing landscapes based on PC framework\label{alg2}}
		\begin{algorithmic}[1] 
			\Require Persistence Diagram $D$ (we assume this is indexed), initial value $m$ and the terminated value $M$, number of mesh points $N$, and level $k$
			\Ensure A vector $V\in\mathbb{R}^{N}$ of values of $\lambda_k(\{m+h\frac{M - m}{N}\}_{h=1}^{N})$
			\State $n\leftarrow\#D$
			\State $B\leftarrow$ $n\times N$ array of Birth Values of $D$ $(B_{i,j} = b_i) $
			\State $E\leftarrow$ $n\times N$ array of Death Values of $D$ ($E_{i,j} = d_i$)
			\State $T\leftarrow$ $n\times N$ array of mesh values ($T_{i,j} = m+j\frac{M - m}{N}$) 
			\State $\Psi\leftarrow$ $n\times N$ array where $\Psi_{i,j} = \min\{(T_{i,j}-B_{i,j})_+,(E_{i,j}-T_{i,j})_+\}$, where $x_+ := \max\{x,0\}$
			\State $\lambda_k\leftarrow$ $N$-dimensional vector where $(\lambda_k)_j = k\max_{1\le i \le n}\{\Psi_{i,j}\}$\\
			\Return $\lambda_k$
		\end{algorithmic}
	\end{algorithm}

{{As we discussed in \Cref{sec:PersistenceCurves}, \Cref{def:persistence curve} is suitable for the numerical implementation. \Cref{alg1} is a pseudo-code for evaluating a generic PC; it is straightforward and consists of three major steps: find $D_t$, the points of $D$ lying in the fundamental box at $t$ for each grid point $t$, evaluate $\psi$ at all points in $D_t$, and finally summarize those values by the operator $T$. Actual implementations are provided by Lawson via the Python package, \texttt{PersistenceCurves} \cite{PCs}.

\Cref{alg1} can be vectorized \footnote{The term ``vectorize'' here means vector operations in a programming language, not to confuse with the vectorization of persistence diagrams.} and take advantage of highly optimized packages, such as \texttt{NumPy} in Python.  In production, the \texttt{PersistenceCurves} package utilizes such vectorized operations, and thus, the codes are shorter and the computations are much faster. As an illustration, \Cref{alg2} is a vectorized algorithm to compute persistence landscapes. As we will see later, \Cref{alg2} is more efficient than the existing persistence landscape in \texttt{GUDHI}. \cite{gudhiPersistenceRepresentations}. 

Computational efforts in \Cref{alg1} are related to $N$, the number of mesh points, and $\#D$, the number of points in a persistence diagram. To test the efficiency of PC computation, we designed two experiments based on different settings of $N$ and $\#D$. We remark that all results in this section were obtained by using Python's \texttt{time} package (and function) on a Lenovo Thinkpad X220 with an Intel I5-2520 CPU and 8G ram. In one experiment, we fixed the number of points in a persistence diagram at $15000$ (for reference, UIUCTex has an average of about 18800 points per diagram) and varied the number of mesh points $N\in\{200n\}_{n=1}^{50}$. To generate the diagrams, we first generated $15000$ birth values from a uniform distribution on $[0,90]$. For each birth value $b$, we randomly sampled a death value from a uniform distribution on $[b,100]$. For each value of $N$, we averaged the computation time from 100 random diagrams times for each method. We performed this experiment to measure the computation times for the normalized life curve ($\mathbf{sl}$), the first persistence landscape computed in the \texttt{PersistenceCurves} package \cite{PCs}, the first persistence landscape computed in \texttt{gudhi}
\cite{gudhiPersistenceRepresentations}\footnote{Here is a direct comparison of two implementations of persistence landscapes. \url{https://gudhi.inria.fr/python/latest/_modules/gudhi/representations/vector_methods.html\#Landscape} and \url{https://github.com/azlawson/PersistenceCurves/blob/master/PersistenceCurves/PC.py\#L76}}, and persistence images computed by \texttt{PersistenceImages}\cite{CSU-TDA}\footnote{For persistence images, the ``number of points in the mesh'' is the number of pixels in the resulting image}.  As shown in \Cref{fig:ComputationTime}(a), we see the computation time increases roughly linearly for each of the functions. The normalized life and life entropy are much faster in computation than the landscapes. We further see a large difference in the \texttt{PersistenceCurves} package landscape computation compared to \texttt{sklearn\_tda}. The difference between the lifespan-based curves and the landscapes can be explained by the fact that the life-based curves do not depend on the input $t$ value in the same way landscapes do. Specifically, The $\psi$ function for $\mathbf{sl}$ and $\mathbf{le}$ does not depend on $t$ while the $\psi$ function for $\lambda_k$ does.

\begin{figure}
    \centering
    \begin{subfigure}{0.45\linewidth}
    \includegraphics[width=\textwidth]{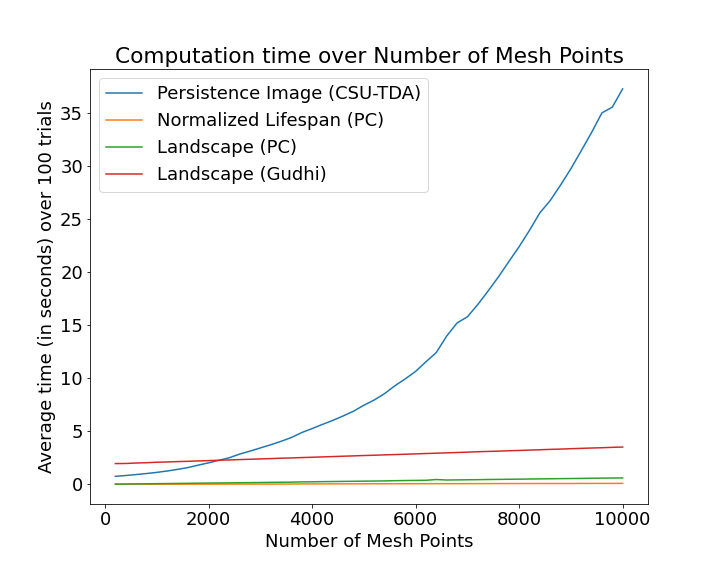}
    \caption{Mesh points}
    \end{subfigure}
    \begin{subfigure}{0.45\linewidth}
    \includegraphics[width=\textwidth]{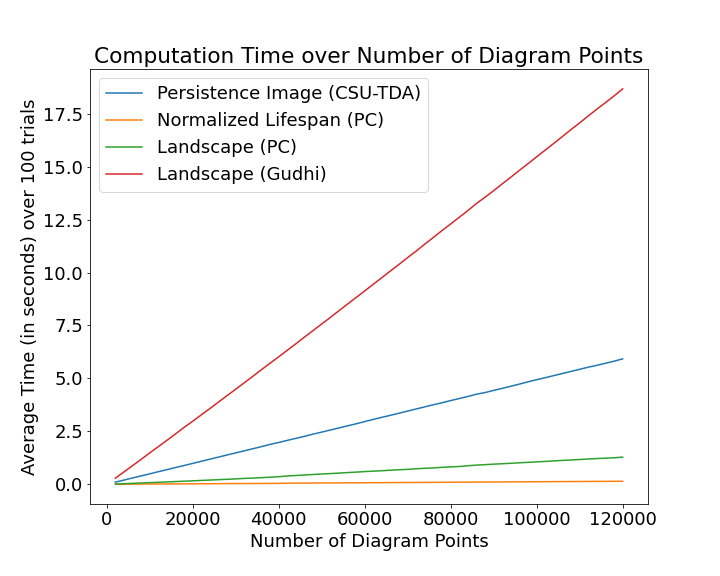}
    \caption{Diagram Points}
    \end{subfigure}
    \caption{Computation time experiments. (a) $\#D$ is fixed and $N$ is varying. (b) $\#D$ is varying and $N$ is fixed. Plots include persistence images computed via the PersistenceImages package from CSU-TDA, the first landscape computed by gudhi, and the normalized lifespan and the first landscape computed by the PersistenceCurves (PC) package.}
    \label{fig:ComputationTime}
\end{figure}

In the second experiment, we measured the computation time of the above mentioned summaries along with persistence images by increasing the number of points in the diagrams from $N \in\{2000n\}_{n=1}^{60}$. Like before, for each value of $N$, we took the average of 100 computation times for each summary. To keep things even, we compute PC methods at 100 points to match the same number of points computed by persistence images (with a resolution of $10\times10$). We see in \Cref{fig:ComputationTime} again the computation time seems linear in the number of diagram points. It is clear both persistence images take significantly and increasingly longer than landscapes, which in turn hold the same relationship to the normalized life curve. 

These experiments show that PCs are very efficient to compute. In fact, normalized life and life entropy never averaged longer than 0.15 seconds per diagram. As stated before, the average number of points in a UIUCTex image is about 18800. At this size, a persistence image takes about 1 second to compute on one diagram with a 10 by 10 resolution. Further considering that UIUCTex contains 1000 images, we see this leads to an estimated 2 seconds per image to generate the model we use in this paper, described in \Cref{sec:applications}. On the other hand, the normalized life curve takes about 0.02 seconds to compute on the same size diagram, leading to an estimated 0.04 seconds per image. Thus we see that the normalized life satisfy the quality of computational efficiency (Quality 3) and indeed all of the newly proposed curves in \Cref{tab:PC} satisfy this quality.}}

\subsection{Applications}\label{sec:applications}
We apply PCs to the following two applications.  The first application, which follows the same experiment as in \cite{adams2017persistence}, is the parameter determination of a dynamical system and the second application is the texture classification. We will introduce the datasets, describe the machine learning models we use, show the numerical results, and compare them with other TDA methods.

\subsubsection{A Discrete Dynamical System}\label{sec:orbits}
\begin{figure}
    \centering
    \includegraphics[width=\textwidth]{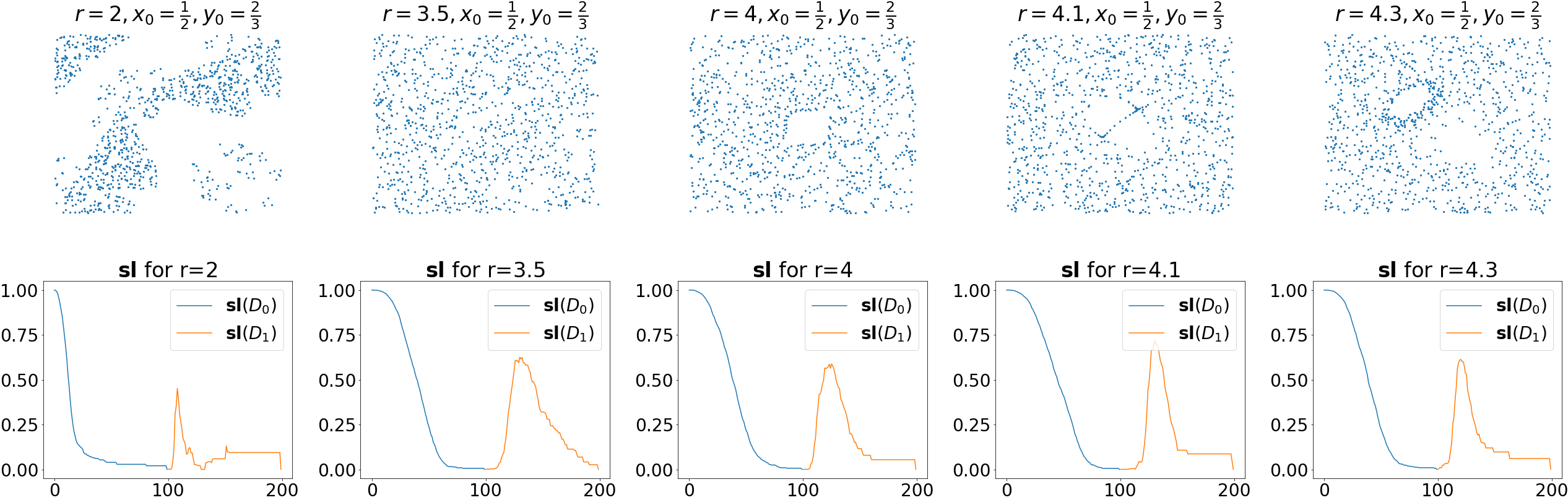}
    \caption{Top panel: examples of linked-twist maps truncated at 1000 points with initial starting position $(\frac{1}{2},\frac{2}{3})$ and $r$ taking on values $2,~3.5,~4,~4.1,~4.3$. Bottom panel: the corresponding normalized lifespan curves.}
    \label{fig:orbits}
\end{figure}

{Adams et. al. \cite{adams2017persistence} proposes a simple experiment that involves generating a dataset by using a linked-twist map, which is a composition of functions  $F, G: [0,1]^2 \to [0,1]^2$. These maps are $F(x, y) = (x + f(y), y)  \hbox{ mod }1$ and $G(x, y) = (x, y+ g(x))  \hbox{ mod }1$ where $f(y) = ry(1-y)$ and $g(x) = rx(1-x)$. In implementation, we discretize this map by selecting initial conditions $x_0$ and $y_0$ along with selecting a parameter $r$. This yields a
discrete dynamical system for modelling fluid flow. The sequence is thus 
}
 \[
 \begin{cases}
 x_{n+1} = x_n + ry_n(1-y_n) &\mathrm{mod} 1\\
 y_{n+1} = y_n + rx_{n+1}(1-x_{n+1}) &\mathrm{mod} 1
 \end{cases}
 \]
 for $n\in\mathbb{N}$. For computations, we must choose to truncate the system at some value. Following \cite{adams2017persistence}, we choose to truncate at 1000 points. The choice of these three parameters lead to a widely varying array of behaviors as indicated by \Cref{fig:orbits} where we fixed the initial position at $(x_0, y_0) = (\frac{1}{2},\frac{2}{3})$ and allowed $r$ to take the values $2,3.5,4,4.1$ and $4.3$. Each plot shows 1000 points.   

We are concerned with determining the parameter $r$ given a set of orbits. For this experiment, we generated 50 orbits for each parameter $r$ each with an initial position that is uniformly sampled from the unit square. We perform one 50/50 train/test split, then trained and scored each model using sci-kit learn's \cite{scikit-learn} \texttt{RandomForest} algorithm with 100 estimators. We repeat the above process 100 times and the reported score for each model is the average of the 100 resulting scores.

Each observation in this dataset is a point cloud in $\mathbb{R}^2$. To compute the persistence diagrams for each point cloud, we use the Vietoris-Rips filtration via \texttt{ripser}\cite{Ripser}. Each of the models below use both 0 and 1-dimensional persistence diagrams, concatenating the resulting vectors for each. The infinite death value for 0-dimensional persistence was replaced with the max death value of the diagram. The models consisted of the following
 \begin{itemize}
 \item[PC:] unaltered, normalized, and entropy versions of the Betti ($\boldsymbol{\beta, \mathbf{s}\beta, \beta \mathbf{e}}$), life ($\mathbf{l,sl,le}$), and midlife ($\mathbf{ml, sml, mle}$) PCs each evaluated at 100 equally spaced points starting at the minimum birth value and ending at the maximum death value generating a 200 dimensional vector for each orbit and each model; 
 \item[$(\lambda_1,\lambda_2,\lambda_3)$:] the first three persistence landscapes $(\lambda_1,\lambda_2,\lambda_3)$ each evaluated at 100 equally spaced points starting at the minimum birth value and ending at the maximum death value generating a 600 dimensional vector for each orbit; 
 \item[ECC:] the Euler Characteristic Curve each evaluated at 100 equally spaced points starting at the minimum birth value and ending at the maximum death value generating a 100 dimensional vector for each orbit; 
 \item[PIM:] Persistence Images with a resolution of $20\times 20$ and $\sigma = 0.005$ (the model proposed in \cite{adams2017persistence}) flattened into a vector generating an 800-dimensional vector for each orbit;
 \item[PS:] Persistence Statistics, which generate a 46 dimensional vector for each orbit. Given a diagram, the statistics are generated from the collections of lifespans $[d-b\mid (b,d)\in  D]$ and midlifes $[\frac{d+b}{2}\mid (b,d)\in  D]$. On these collections, we calculate the mean, standard deviation, skewness, kurtosis, 10th, 25th, 50th, 75th, 90th percentiles, IQR, and coefficient of variation. Finally, we calculate the persistent entropy \cite{persistentEntropy} of diagrams.
 \end{itemize}

 The reported scores appear in the last column of \Cref{tab:Scores}. The upper half of the table lists scores of the individual models described above, while the lower half list scores of the models above concatenated with persistence statistics. For this experiment, in both halves, we see that the PCs proposed in this paper, indicated by (PC) perform well. 

 In the upper half of the \Cref{tab:Scores}, we see $\mathbf{sl}$ performs the best. The life entropy curve (persistent entropy summary) \cite{persistentEntropyStability} is the only model not originally proposed in this work to appear in the top 5.  In the lower half of the table, we concatenated the original models with persistence statistics. This additional information seems to boost the classification power of the original models in most cases.  With the addition of the persistence statistics, we see that the top 5 models are nearly equivalent.

 \subsubsection{Texture Classification}
 Texture classification is a classic task in the field of computer vision. In this paper, we apply our models to three texture databases.  
 \begin{itemize}
     \item [\textbf{Outex0}:] A database from the University of Oulu and consists of $15$ test suites each with a different challenge \cite{outex2002}.  We focus on the test suites $0$, which contains $24$ texture classes with $20$ grayscale images of each class that are $128$ by $128$ in size. The suite is equipped with $100$ preset 50/50 train/test splits. A score on the test suite is the average accuracy over all $100$ pre-defined splits.
     \item [\textbf{UIUCTex}:] A texture database from University of Illinois at Urbana-Champaign \cite{lazebnik2005sparse}. The dataset consists of $25$ texture classes with $40$ grayscale images of each texture. Each image is of size 480 by $640$. Following the methods of \cite{perea2014klein}, a score on this set is the average accuracy of 100 random 80/20 train/test splits.
     \item [\textbf{KTH:}] KTH's Textures under varying Illumination, Pose and Scale (KTH-TIPS2b) is a database containing 81 200 by 200 grayscale images for each of its 10 textures \cite{KTHTIPS2}. As the name suggests, each texture class contains images of different scales, rotations, and illuminations. A score on this set is the average accuracy of 100 random 80/20 train/test splits. 
 \end{itemize}

\begin{figure}

    \centering
    \begin{subfigure}{0.3\textwidth}
        \centering
        \includegraphics[width=\textwidth]{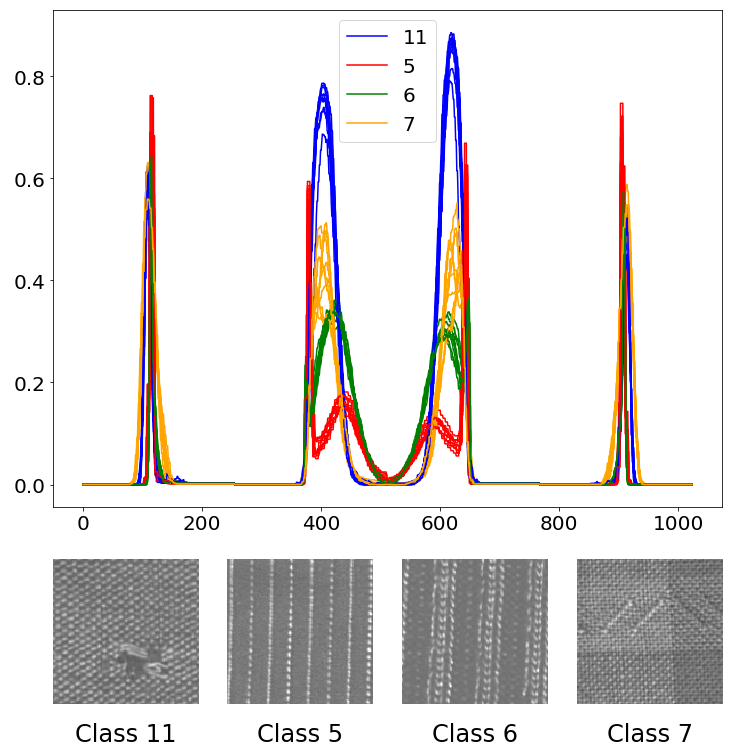}
        \caption{Outex}\label{PCs}
    \end{subfigure}
    \begin{subfigure}{0.3\textwidth}
        \centering
        \includegraphics[width=\textwidth]{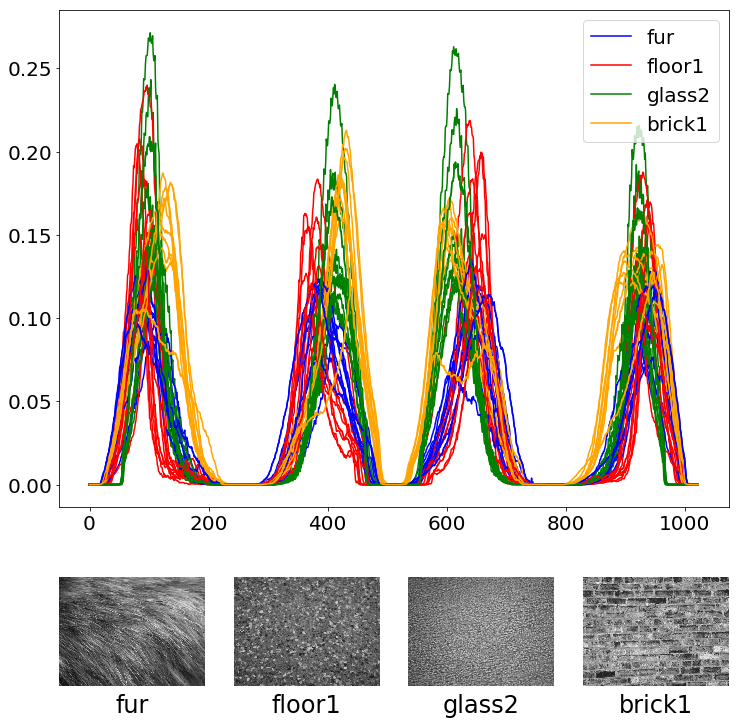}
        \caption{UIUCTex}
        \label{fig:UIUCCurves}
    \end{subfigure}
    \begin{subfigure}{0.3\textwidth}
        \centering
        \includegraphics[width=\textwidth]{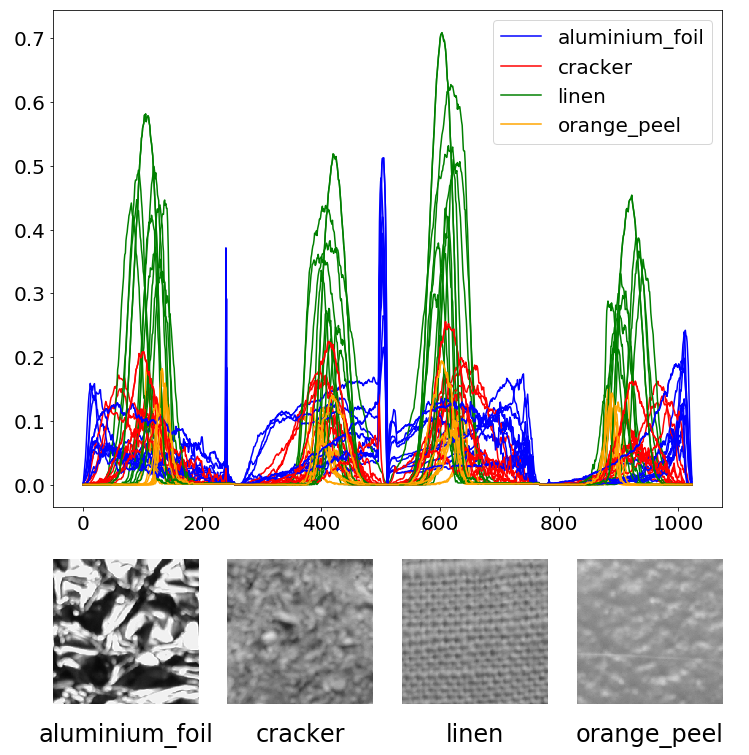}
        \caption{KTH-TIPS}
         \label{fig:KTHCurves}
    \end{subfigure}

    \caption{The PCs ($[\mathbf{ml}_0(G),\mathbf{ml}_1(G),\mathbf{ml}_0(G^C),\mathbf{ml}_1(G^C)]$) for each image $G$ in 3 selected classes from the specified database, where $G^C$ represents the complement image of $G$.}\label{fig:dbPCsamples}

\end{figure}

The models use the same curves and methods as used in \Cref{sec:orbits} with a few notable differences.
Each observation for each database is an 8-bit grayscale image, which is a 2D array whose entries take integer values between 0 and 255.  Persistence diagrams of an 8-bit image are calculated based on cubical complex and sublevel set filtration. Images have a guaranteed minimum birth of 0 and maximum death of 255. Moreover, all critical values are guaranteed to occur at integer values. Hence, for each diagram, all PC-based methods use the integers from $0$ to $255$ to generate a 256-dimensional vector.  Moreover, as discussed in \cite{chung2018topological}, computing persistence diagrams on images is affected by the boundary effect. Briefly, the boundary effect causes potential $H_1$ generators lying on the boundary of an image to go uncounted. To account for this, we also consider the persistent homology of the complement (or inverse) of an image. In total, each model considers four diagrams: the 0 and 1-dimensional diagrams for an image and its inverse. \Cref{fig:dbPCsamples} shows sample images from three texture database and their persistence curves.  This leads to a 3072-dimensional vector for the first three landscapes model, 512-dimensional vector for the ECC method, a 92-dimensional vector for persistence statistics, a 1600-dimensional vector for persistence images, and 1024-dimensional vectors for all other PC-based methods. Finally, for persistence images, we set $\sigma = 1$.

The first three columns of \Cref{tab:Scores} displays our reported scores for the models on these three databases. In the upper half of the \Cref{tab:Scores}, we see that the normalized lifespan curve is among the top 5 performers on KTH-TIPS2b, coming in second behind persistence statistics and UIUCTex, coming in third. Persistence images performs well across all three texture datasets and performs best on UIUCTex. Perhaps most surprisingly is the performance of the standalone persistence statistics, which uses only 92-dimensional vectors to achieve the best score on both KTH-TIPS2b and Outex. Both the Euler Characteristic Curve and persistence landscapes fail to place here.

{
\begin{remark}
It is interesting to compare the performance of persistence landscapes, which enjoy a high level of stability, with that of the Betti curve or persistent statistics, which do not. The numerical evidence here suggests that stability and classification performance might not be highly correlated.
\end{remark}
}

In the lower half of the \Cref{tab:Scores}, we see again that persistence statistics improve the classification power of the original models. Top performers include normalized life, taking second and third on KTH-TIPS2b and UIUCTex respectively and persistence images, which was the top performer on KTH-TIPS2b and UIUCTex. Here, we see that the persistence landscapes model does not make it into the top 5, but the Euler Characteristic Curve model does on Outex and KTH-TIPS2b.

There are other TDA methods with applications to the same datasets. Outex0 has been studied through Sparse-TDA~\cite{guo2018sparse}, persistence scale space kernel~\cite{reininghaus2015stable}, metric learning for persistence diagrams~\cite{li2014persistence}, sliced Wasserstein Kernel~\cite{SlicedWassersteinKernel}, and persistence paths (specifically, the Betti kernel) \cite{chevyrev2018persistence}. The reported classification scores from those work are $66.0$, $69.2$, $87.5$, $98.8$, and $97.8$ respectively.  In \cite{perea2014klein}, authors applied their methods to both UIUCTex and KTH-TIPS2b, and their reported classification scores are $91.2$, and $94.8$, respectively.  As a reference, the state-of-arts classification scores based on traditional bag of words based texture representation for these datasets are $99.5$ \cite{feichtinger2012gabor} for Outex0, $99.0$ \cite{liu2019bow} for UIUCTex, and $99.4$ \cite{liu2019bow} for KTH-TIPS2b.

\begin{table}
    \centering
    \begin{tabular}{|c|c|c|c||c|}\hline
    & Outex0& UIUCTex & KTH-TIPS2b& Orbits\\\hline
PS \cite{chung2018topological}& \cellcolor{green!20}$\mathbf{98.32\pm0.89}$&$90.51\pm2.06$ & \cellcolor{green!20} $\mathbf{95.47\pm1.55}$& $86.74  \pm  3.00$\\\hline
$\boldsymbol{\beta}$&\cellcolor{green!20} $97.31 \pm1.10$ & $88.71 \pm 2.26$ & $90.92\pm 1.92$&$88.20  \pm  2.97$\\\hline

s$\boldsymbol{\beta}$(PC)  &\cellcolor{green!20}$97.55\pm 1.11$&$91.71 \pm2.03$& $91.01\pm 2.07$&$88.46  \pm  2.89$\\\hline

$\boldsymbol{\beta}$e(PC) &\cellcolor{green!20}$97.47\pm 1.14$& \cellcolor{green!20}$91.98\pm 1.99$&\cellcolor{green!20}$91.38\pm 1.91$ &$88.02  \pm  2.92$\\\hline

$\mathbf{l}$ (PC)& $95.75 \pm1.40$&$88.67\pm 2.16$& $90.73\pm 2.36$&\cellcolor{green!20}$90.11  \pm  2.47$\\\hline

$\mathbf{sl}$ (PC)&$96.83\pm 1.22$& \cellcolor{green!20}$92.75\pm 1.78$&\cellcolor{green!20}$93.64\pm 2.24$ &\cellcolor{green!20}$\mathbf{90.13  \pm  2.50}$\\\hline

$\mathbf{le}$ \cite{persistentEntropyStability} &$96.76\pm 1.21$&\cellcolor{green!20}$92.82 \pm1.65$& $91.12\pm 2.11$&\cellcolor{green!20}$89.54  \pm  2.49$\\\hline

$\mathbf{ml}$  (PC)&$97.26\pm 1.16$&$89.00\pm 2.13$& $90.68\pm 2.11$&$89.11  \pm  2.91$\\\hline

$\mathbf{sml}$  (PC)&$97.29\pm 1.05$&$91.83 \pm1.79$&\cellcolor{green!20}$91.54\pm 2.17$ &\cellcolor{green!20}$89.88  \pm  2.57$\\\hline

$\mathbf{mle}$  (PC)& $97.28\pm 1.00$&\cellcolor{green!20}$92.10 \pm1.83$&$91.16\pm 1.96$&\cellcolor{green!20}$89.55  \pm  2.47$\\\hline

Land\cite{bubenik2015statistical} &$90.0\pm1.65$ & $87.92\pm 2.19$&$87.68\pm2.38$&$89.51  \pm  2.74$\\\hline

ECC &$96.10\pm1.24$ &$85.08\pm2.12$& $89.17\pm2.31$& $71.42  \pm  3.47$\\\hline

PIM\cite{adams2017persistence} &\cellcolor{green!20}$97.65\pm 0.91$ &\cellcolor{green!20}$\mathbf{94.52\pm 1.42}$ & \cellcolor{green!20}$93.40 \pm1.69$&$84.86  \pm  3.09$ \\\hline
With PS\\\hline

$\boldsymbol{\beta}$+PS& \cellcolor{green!20}$\mathbf{98.71\pm 0.82}$& $91.30\pm 1.92$& \cellcolor{green!20}$94.86\pm 1.65$&$89.27  \pm  2.90$\\\hline

s$\boldsymbol{\beta}$+PS&$98.27 \pm0.82$ & \cellcolor{green!20}$93.10 \pm1.81$&$94.35 \pm1.60$ &$89.29  \pm  2.80$\\\hline

$\boldsymbol{\beta}$e+PS&$98.18\pm 0.86$ &\cellcolor{green!20} $93.16 \pm1.75$&$94.72\pm 1.73$ &$89.13  \pm  2.88$\\\hline

$\mathbf{l}$+PS&$98.27\pm 0.91$ & $92.85 \pm1.81$& \cellcolor{green!20}$95.22\pm 1.80$ &\cellcolor{green!20}$90.37  \pm  2.54$\\\hline

$\mathbf{sl}$+PS&$97.95\pm 0.97$ &\cellcolor{green!20}$ 94.25 \pm1.48$&\cellcolor{green!20}$95.54\pm 1.55$ &\cellcolor{green!20}$90.42  \pm  2.43$\\\hline

$\mathbf{le}$+PS&$98.05\pm 1.01$&\cellcolor{green!20}$  94.32 \pm1.39$&$94.81\pm 1.75$&\cellcolor{green!20}$90.18  \pm  2.49$\\\hline

$\mathbf{ml}$+PS&\cellcolor{green!20}$98.60\pm 0.937$&$91.68 \pm1.90$&$94.85\pm 1.59$&$89.47 \pm   2.83$\\\hline

$\mathbf{sml}$+PS&\cellcolor{green!20}$98.07\pm 0.87$&$  92.68\pm 1.96$&$94.72\pm 1.86$ &\cellcolor{green!20}$\mathbf{90.50  \pm  2.63}$\\\hline

$\mathbf{mle}$+PS&$98.05\pm 0.90$& $92.68 \pm1.51$&$94.61 \pm1.73$&$90.12  \pm  2.63$\\\hline

Land+PS &$96,58\pm 1.20$ & $92.35\pm 1.85$&$93.41\pm2.10$ &\cellcolor{green!20}$90.23  \pm  2.56$\\\hline

ECC+PS & \cellcolor{green!20}$98.60\pm0.85$ & $92.27\pm1.58$& \cellcolor{green!20}$94.93\pm1.92$& $84.51  \pm  3.43$\\\hline

PIM+PS & \cellcolor{green!20}$98.66\pm0.71$ &\cellcolor{green!20}$\mathbf{94.87\pm 1.53}$ &\cellcolor{green!20}$\mathbf{95.83\pm 1.56}$ & $87.27 \pm   3.10$\\\hline
    \end{tabular}
    \caption{Performances on Outex, UIUCTex, KTH-TIPS2b, and Orbits. The table is split into two parts: the top part lists each individual model, while the bottom part displays the top models concatenated with persistence statistics. In each part, the largest score for each column is bold and the cells for the top 5 scores in each column is shaded light green.}
    \label{tab:Scores}
\end{table}

\subsection{Noise Handling}
In order to demonstrate how each vectorization reacts to the noise, we
design two experiments using the Outex0 and KTH-TIPS2b databases. We compare the performance of the normalized life curve, life entropy curve, first three landscapes, persistence statistics, and persistence images models by adding Gaussian noise to the testing set for the scoring process of each database. The noise was added with a standard deviation fixed at values in $\{0,~0.5,~1,~1.5,~2,~2.5,~3,~3.5,~4,~4.5,~5\}$. The results are plotted in \Cref{fig:CurveLimitation}(a)-(b). For Outex0, we see that persistence images are most resistant to noise and persistence statistics are least resistant. The normalized life, life entropy, and landscape curves perform similarly and begin to fall away from persistence images around $\sigma=2.5$. It seems in \Cref{fig:CurveLimitation}(a) that these curves are leveling to a linear loss of score while the persistence image model seems to increase its loss over $\sigma$. In the KTH-TIPS2b experiment, we see a similar result as shown in \Cref{fig:CurveLimitation}(b). Persistence images seem to be the most stable followed closely by normalized life and life entropy. Persistence statistics falls off quickly as do persistence landscapes.  Because the images in KTH-TIPS2b are larger, the noise has less effect leading to more stability.  { This experiment demonstrates empirically how  each vectorization reacts to noise under the random forest classification model.  It would be interesting to investigate rigorously the relation between the performances of different classification models and stability of the vectorizations. }

Finally we offer advice to practitioners that choose to use PCs. PCs carry interpretable information about the diagrams they are calculated from. In turn, the diagrams carry interpretable information about the topological properties of the underlying space. This allows domain experts to reasonably choose PCs to use in practice. However, the need for a learnable method is not far from our current setup. For example, take a general $\psi$ function as  $\psi(D,b,d,t) = A\times b + B\times d +C$ where $A,~B,~C\in\mathbb{R}$. If $T$ is the sum statistic, then the resulting PC is one from which we can derive the life ($A=-1,~B=1,~C=0)$, midlife ($A=0.5,~B=0.5,~C=0$), and Betti ($A=0,~B=0,~C=1$) curves. We now have a class of PCs controlled by the parameters $A,~B,~C$. Immediately, then, we may use a grid-search algorithm to select these parameters and learn the best curve for a given problem. We further note that both normalized and entropy-based curves are simple modifications of these basic curves, and hence fall under this grid-search framework as well. A deeper exploration into this facet of PCs will be left for future work. 
\subsection{Limitations and Considerations}
\begin{figure}
\centering
\begin{subfigure}{0.33\linewidth}
\begin{center}
    \includegraphics[width=\textwidth]{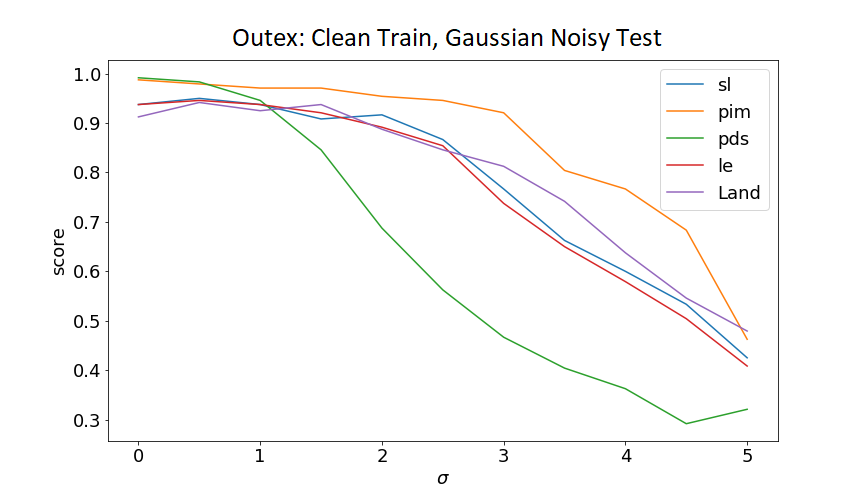}
\end{center}
\caption{Outex}
\end{subfigure}
~
\begin{subfigure}{0.33\linewidth}
\begin{center}
    \includegraphics[width=\textwidth]{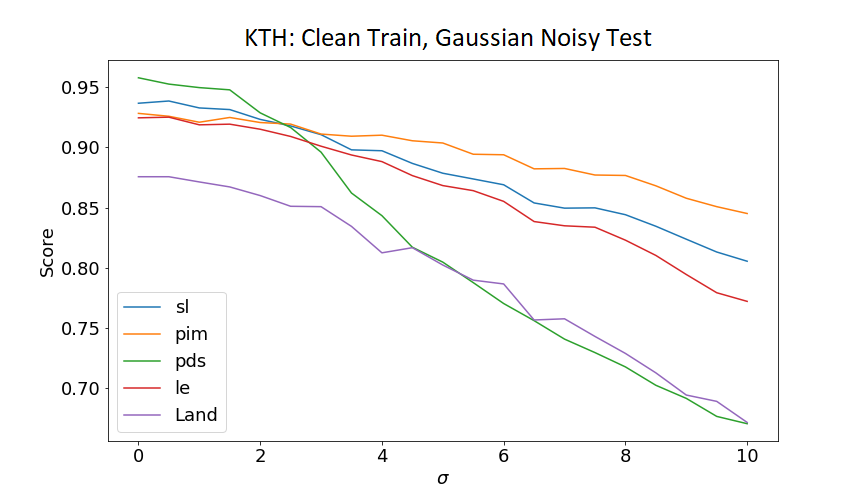}
\end{center}
\caption{KTH}
\end{subfigure}
\begin{subfigure}{0.14\linewidth}
\begin{center}
    \includegraphics[width=\textwidth]{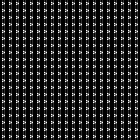}
\end{center}
\caption{~}
\end{subfigure}
~
\begin{subfigure}{0.14\linewidth}
\begin{center}
    \includegraphics[width=\textwidth]{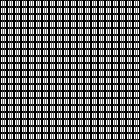}
\end{center}
\caption{~}
\end{subfigure}
    \caption{(a)-(b) Noise handling experiments. Scores on Outex0 and KTH-TIPS2b obtained by training on a clean training set and adding Gaussian noise with mean $0$, and variance $\sigma^2$ to images in the testing set. (c)-(d) Limitations. These two figures seem to have different textures, but they yield the same persistence diagrams (not shown).}
    \label{fig:CurveLimitation}
\end{figure}
As with any method, PCs come with some limitations. First and foremost, the information they carry about the original space is limited by the amount of information carried by the persistent homology. For example, the two images displayed in \Cref{fig:CurveLimitation}(c)-(d) appear to have different patterns and yet they yield the same persistence diagrams. This leads to equal PCs and hence the images are indistinguishable via the normal persistence process. It can be useful to consider the original space in addition to the topology in the analysis, and model construction. 

When working with PCs, or any vectorization method, one must also keep in mind the mesh (or resolution) of the vector. For image analysis, the choice of integer values from 0 to 255 is a natural one; however, in randomly generated point-cloud data, the choice is not so clear. The choice becomes less clear when dealing with a more dynamic dataset, such as one that does not have a minimum birth nor a maximum death value. Balancing the vector size, with the information retained is important to keep in mind. Finally, based on our experiments, we would recommend that one choose the normalized life curve $\mathbf{sl}$ among all other curves presented here for its stability, computational efficiency, and performance.

\section{Generalization and Conclusion}
\label{sec:conclusion}
PCs provide a simple general framework for generating functional summaries and vectorizations of persistence diagrams. These curves are compatible with machine learning algorithms, they can be stable. They are efficient to compute, and by choice of functions and statistics, one can alter the importance of points in different regions of the persistence diagrams. They also perform well in real applications as seen through the classification experiments for textures and the discrete dynamical system parameters. The theory and experimentation presented here are by no means complete. We conclude this paper by listing below several potential directions for further investigation. 
\begin{questions}$\;$

\begin{itemize}

    \item[Q1] In \Cref{thm:main stability thm}, the operator $T$ is fixed as $\Sigma$.  What conditions on the function $\psi$ or the statistic $T$ can lead to a more general and useful stability result?
    \item[Q2]  Several statistical properties, such as laws of large numbers, and stochastic convergence \cite{chazal2015subsampling,chazal2014stochastic,bubenik2018persistence}, of persistence landscapes has been established.  Since persistence landscapes fall under the PC framework, it would be interesting to investigate general conditions on $\psi$ and $T$ so that the same properties hold.
    \item[Q3] The Euler Characteristic Transform \cite{turner2014persistent} was proved to be a sufficient statistic for distributions on the space of subsets of $\mathbb{R}^d$ that can be written as simplicial complexes where $d=2,~3$, and was applied to shape analysis.  It would be interesting to generalize this concept to PCs. 
    \item[Q4] Is there a statistical framework to perform ``curve selection'' that will produce an optimal or near optimal set of curves for modeling?
    \item[Q5] Since \Cref{lem:estimate} can be generalized to the $L^p$-norm, it would be interesting to investigate the stability results with respect to $W_p$ distance.

\end{itemize}
\end{questions}

\bibliographystyle{siam}
\bibliography{main}

\appendix
\section{Explicit Calculations of Bounds Using \Cref{thm:main stability thm} }\label{app:stability}
In this section, we will assume $C,D\in\mathcal{D}$ and utilize the definitions of the curves defined in \Cref{tab:PC}. To ease notation we will write $\kappa_1$ in place of $\kappa_1(\psi,C,D)$ and analogously for $\kappa_\infty,\delta_1,\delta_\infty$. We recall the following definitions. 
\begin{align}
 \kappa_{1}(\psi, C, D) & =\sum_{(b,d)\in C\setminus\Delta}\max_{ t\in [b, d]} |\psi(b, d, t)| +  \sum_{(b'd')\in D\setminus\Delta}\max_{t\in [b', d']} |\psi(b',d', t)|\\
 \kappa_{\infty}(\psi, C, D)  &=\max_{(b,d)\in C\setminus\Delta}\max_{ t\in [b, d]} |\psi(b, d, t)| +  \max_{(b',d')\in D\setminus\Delta}\max_{t\in [b', d']} |\psi(b', d', t)|.\\
\delta_{1}(\psi, C, D) &=\inf_{\eta:C\to D}\sum_{i=1}^{n_\eta}\max_{{ t\in [b_i, d_i]\cap [\eta_{b_i}, \eta_{d_i}]}} |\psi(b_i, d_i, t) - \psi(\eta_{b_i}, \eta_{d_i}, t)|.\\
\delta_{\infty}(\psi, C, D) &=\inf_{\eta:C\to D}\max_{\substack{1\leq i \leq n_\eta \\ t\in [b_i, d_i]\cap [\eta_{b_i}, \eta_{d_i}]}} |\psi(b_i, d_i, t) - \psi(\eta_{b_i}, \eta_{d_i}, t)|.
\end{align}

\subsection{Betti-based Curves}
In this section we consider the curves based on the Betti number. As we will see the $\psi$ function for each of the curves in this section is not continuous at the diagonal, which causes the $\delta$ bounds to be large. 
\subsubsection{Betti Number Curve ($\boldsymbol{\beta}$)}
Recall that the Betti curve is by taking $\psi(D;b,d,t) :=\psi^D(b,d)= 1$ when $(b,d)\in D$, $b\neq d$ and 0 otherwise. The values for $\kappa_1$ and $\kappa_\infty$ are straightforward to calculate.
\begin{align}
    \kappa_1 &= \sum_{i=1}^{n^c}1 + \sum_{i=1}^{n^D} 1  =n^C + n^D,\nonumber\\
    \kappa_\infty &= 2.\nonumber
\end{align}
On the other hand, notice that if $t\in[b,d]\cap[\eta_b,\eta_d]$ then we have $\psi^C(b,d) - \psi^D(\eta_b,\eta_d) = 1$ when $(b,d)\in\Delta \Rightarrow (\eta_b,\eta_d)\notin\Delta$ and 0 otherwise. This means that the worst case scenario for the Betti curve is when all points are paired with the diagonal yielding the following $\delta$ values
\begin{align}
    \delta_\infty &\le 1.\nonumber
\end{align}
\begin{align}
    \delta_1 &\le (n^C+ n^D).\nonumber
\end{align}
Hence, we may conclude by \Cref{thm:main stability thm} that
\begin{align}
    \|\beta(C)-\beta(D)\|_1&\le (n^C+ n^D)W_\infty(C,D) + L^C\land L^D, \label{equ:betti Winfty bound}\\
    \|\beta(C)-\beta(D)\|_1&\le 2W_1(C,D) + (L_\infty^C\land L_\infty^D)(n^C+ n^D). \label{equ:betti W1 bound}
\end{align}
Neither of these bounds are desirable. Not only as the number of points increase both bounds tend to infinity, but also both bounds contain constants that are irrelevant to $W_1$ nor $W_{\infty}$. Moreover, the same occurs as the minimum lifespan grows.

\subsubsection{Normalized Betti Curve ($\mathbf{s}\boldsymbol{\beta}$)}
According to \Cref{def:normalized curve}, we may define the normalized Betti curve by taking $\psi(D;b,d,t) :=\psi^D(b,d)= \frac{1}{n^D}$ when $(b,d)\in D$, $b\neq d$ and $0$ otherwise. Again we can get the $\kappa$ values quickly
\begin{align}
    \kappa_1 &=\sum_{i=1}^{n^C} \frac{1}{n^C} + \sum_{i=1}^{n^D}\frac{1}{n^D}=2,\nonumber\\
    \kappa_\infty &= \frac{1}{n^C}+\frac{1}{n^D}\le\frac{2}{n^C\land n^D},\nonumber
\end{align}
On the other hand, for all $\eta:C\to D$
\begin{align}
    \delta_\infty &\le \max_{1\le i\le n_\eta}|\psi^C(b_i,d_i)-\psi^D(\eta_{b_i},\eta_{d_i})|\le \frac{1}{n^C\land n^D}.\nonumber\\
    \delta_1 &\le \sum_{i=1}^{n_\eta}|\psi^C(b_i,d_i)-\psi^D(\eta_{b_i},\eta_{d_i})| \le \sum_{i=1}^{n^C} \frac{1}{n^C} + \sum_{i=1}^{n^D}\frac{1}{n^D}=2.\nonumber
    \end{align}
Hence, we may conclude by \Cref{thm:main stability thm} that
\begin{align}
    \|\mathbf{s}\boldsymbol\beta(C)-\mathbf{s}\boldsymbol\beta(D)\|_1&\le 2W_\infty(C,D) + \frac{L^C\land L^D}{n^C\land n^D},\label{equ:normalized betti Winfty bound}\\
    \|\mathbf{s}\boldsymbol{\beta}(C)-\mathbf{s}\boldsymbol{\beta}(D)\|_1&\le \frac{2W_1(C,D)}{n^C\land n^D} + 2(L_\infty^C\land L_\infty^D). \label{equ:normalized betti W1 bound}
\end{align}
Even with the normalization, these two bounds still depend heavily on the number of diagram points and the maximum lifespan.

\subsubsection{Betti Entropy Curve ($\boldsymbol{\beta}\mathbf{e}$)}
By \Cref{def:entropy-based persistence curve}, the Betti Entropy Curve is defined by taking $\psi(D;b,d,t)=\psi^D(b,d) = -\frac{1}{n^D}\log\frac{1}{n^D}$ when $(b,d)\in D, b\neq d$ and 0 otherwise.  If the points of $C$ and $D$ are indexed according to the optimal matching for $W_\infty(C,D)$ distance and if $n^C\land n^D \ge 3$ (so that $\frac{1}{n^C\land n^D}\le \frac{1}{e}$), then by \Cref{lem:loglemma}
\begin{align}
    \kappa_1 &= \sum_{i=1}^n \psi^C(b_i,d_i) + \sum_{i=1}^n \psi^D(b_i,d_i)  =-\log \frac{1}{n^C}-\log \frac{1}{n^D} \le 2\log(n^C\lor n^D),\nonumber\\
    \kappa_\infty &= \max_{1\le i \le n}\psi^C(b_i,d_i) + \max_{1\le i \le n}\psi^D(b_i,d_i) \le\frac{2}{e},\nonumber
\end{align}

On the other hand, if the points of $C$ and $D$ are indexed according to the optimal matching for $W_1(C,D)$, then by \Cref{lem:loglemma}
\begin{align}
    \delta_\infty &\le \max_{1\le i\le 1}|\psi^C(b_i,d_i)-\psi^D(\eta_{b_i},\eta_{d_i})|\le -\frac{1}{n^C\land n^D}\log \frac{1}{n^C\land n^D}.\nonumber\\
     \delta_1 &\le \sum_{i=1}^n|\psi^C(b_i,d_i)-\psi^D(\eta_{b_i},\eta_{d_i})|\le -\frac{n^C\lor n^D}{n^C\land n^D}\log \frac{1}{n^C\land n^D}. \nonumber
    \end{align}
Hence we may conclude by \Cref{thm:main stability thm} that if $n^C\land n^D\ge 3$ then
\begin{align}
    \|\boldsymbol\beta\mathbf{e}^C-\boldsymbol\beta\mathbf{e}^D\|_1&\le 2\log(n^C\lor n^D)W_\infty(C,D) - (L^C\land L^D)(\frac{1}{n^C\land n^D}\log \frac{1}{n^C\land n^D}),\label{equ:betti entropy Winfty bound}\\
    \|\boldsymbol{\beta}\mathbf{e}^C-\boldsymbol{\beta}\mathbf{e}^D\|_1&\le \frac{2W_1(C,D)}{e} + (L_\infty^C\land L_\infty^D)\frac{n^C\lor n^D}{n^C\land n^D}\log \frac{1}{n^C\land n^D}. \label{equ:betti entropy W1 bound}
\end{align}

\subsection{Midlife-based Curves}
For this section, we will assume that all birth and death values are non-negative (so that $b+d\neq 0$). The midlife-based curves, similar to the lifespan based curves, benefit from continuous $\psi$ functions. Moreover, the bounds in this section take a form similar to those of the lifespan-based counterparts. For this section, we will use the notation $u_i^D:=\frac{\eta_{b_i}+\eta_{d_i}}{2}$, $U^D:=\sum_{i=1}^{n^D}u_i^D$, and $U_\infty^D := \max_{i}u_i^D$. Finally, because we are assuming non-negativity, we also have $2U^D \ge L^D$.
\subsubsection{Midlife Curve $\mathbf{ml}(D)$}
The midlife curve take $\psi(D;,b,d,t) := \psi^D(b,d) = \frac{b+d}{2}$. The values $\kappa_1 = U^C+U^D\le 2(U^C\lor U^D)$ and $\kappa_\infty = U_\infty^C + U_\infty^D\le U_\infty^C \lor U_\infty^D$. We see then that for any $\eta:C\to D$
\begin{align*}
    \delta_\infty &\le \max_{1\le i\le n}|\psi^C(b_i,d_i)-\psi^D(\eta_{b_i},\eta_{d_i})|\\
    &=\max_{1\le i\le n}\left|\frac{(d_i+b_i)}{2}-\frac{(\eta_{d_i}+\eta_{b_i})}{2}\right|\\
    &\le\frac{1}{2}\max_{1\le i\le n}|(d_i-\eta_{d_i})|+|(b_i-\eta_{b_i})|\\
    &\le W_\infty(C,D).
\end{align*}
If instead the indexing follows the optimal matching for $W_1(C,D)$ we see 
\begin{align*}
    \delta_1 &\le \sum_{i=1}^{n}|\psi^C(b_i,d_i)-\psi^D(\eta_{b_i},\eta_{d_i})|\\
    &=\sum_{i=1}^{n}\left|\frac{(d_i+b_i)}{2}-\frac{(\eta_{d_i}+\eta_{b_i})}{2}\right|\\
    &\le\frac{1}{2}\sum_{i=1}^{n}|(d_i-\eta_{d_i})|+|(b_i-\eta_{b_i})|\\
    &\le W_1(C,D).
\end{align*}
Therefore by \Cref{thm:main stability thm} we conclude
\begin{align}
    \|\mathbf{ml}(C)-\mathbf{ml}(D)\|_1&\le 2(U^C\lor U^D)W_\infty(C,D) + (L^C\land L^D)W_\infty(C,D), \label{equ:midlife Winfty bound}\\
    \|\mathbf{ml}(C)-\mathbf{ml}(D)\|_1&\le 2(U_\infty^C\lor U_\infty^D)W_1(C,D) + (L_\infty^C\land L_\infty^D)W_1(C,D). \label{equ:midlife W1 bound}
\end{align}

\subsubsection{Normalized Midlife Curve $\mathbf{sml}(D)$} 
The normalized midlife curve take \\ $\psi(D;,b,d,t) := \psi^D(b,d) = \frac{b+d}{{2U^D}}$. The values $\kappa_1 \le 2$ and $\kappa_\infty \le 2$. Moreover, suppose $L^C\le L^D$. We note that $|U^C-U^D|\le\sum_{i=1}^n|u_i^C-u_i^D|\le nW_\infty(C,D)$.
\begin{align*}
    \delta_\infty &\le \max_{1\le i\le n}|\psi^C(b_i,d_i)-\psi^D(\eta_{b_i},\eta_{d_i})|\\
    &\le\max_{1\le i\le n}\frac{|u_i^C-u_i^D|}{U^D} +\max_{1\le i\le n}u_i^C\frac{|U^C-U^D|}{U^CU^D} \\
    &\le\max_{1\le i\le n}\frac{|b_i+d_i-\eta_{b_i}-\eta_{d_i}|}{L^D} +\max_{1\le i\le n}\frac{4(U^C_\infty\lor U^D_\infty)|U^C-U^D|}{L^CL^D}\\
    &\le \frac{2W_\infty(C,D)}{L^C\lor L^D} + \frac{4n(U^C_\infty\lor U^D_\infty)W_\infty(C,D)}{(L^C\land L^D)(L^C\lor L^D)}. 
\end{align*}
Moreover,
\begin{align*}
    \delta_1 &\le \sum_{i=1}^{n}|\psi^C(b_i,d_i)-\psi^D(\eta_{b_i},\eta_{d_i})|\\
    &\le\sum_{i=1}^n\frac{|u_i^C-u_i^D|}{U^D} +\sum_{i=1}^nu_i^C\frac{|U^C-U^D|}{U^CU^D} \\
    &\le\sum_{i=1}^n\frac{|b_i+d_i-\eta_{b_i}-\eta_{d_i}|}{L^D} +\sum_{i=1}^n\frac{2|U^C-U^D|}{L^D}\\
    &\le \frac{4W_1(C,D)}{L^C\lor L^D}.
\end{align*}
Therefore by \Cref{thm:main stability thm} we conclude
\begin{align}
    \|\mathbf{sml}(C)-\mathbf{sml}(D)\|_1&\le 2W_\infty(C,D) + (L^C\land L^D)(\frac{2W_\infty(C,D)}{L^C\lor L^D} + \frac{4n(U^C_\infty\lor U^D_\infty) W_\infty(C,D)}{(L^C\land L^D)(L^C\lor L^D)}), \nonumber\\
    &\le4W_\infty(C,D)(1+\frac{U_\infty^C\lor U_\infty^D}{L^C\lor L^D}n) \label{equ:normalized midlife Winfty bound} \\
    \|\mathbf{sml}(C)-\mathbf{sml}(D)\|_1&\le 2W_1(C,D) + L_\infty^C\land L_\infty^D\frac{4W_1(C,D)}{L^C\lor L^D} \le 6W_1(C,D).\label{equ:normalized midlife W1 bound}
\end{align}

\subsubsection{Midlife Entropy Curve $\mathbf{mle}(D)$}
In this final section we will discuss the bounds for the midlife entropy curve, which uses $\psi(D;b,d,t):=\psi^D(b,d)=-\frac{b+d}{U^D}\log\frac{b+d}{U^D}$. Straightforward calculation reveals $\kappa_1 = \log n^C + \log n^D$ and $\kappa_\infty\le \frac{2}{e}$. The bounds recovered for midlife entropy are the same as for life entropy. That is, if $r_\infty(C,D),r_1(C,D)\le \frac{1}{2e}$, \Cref{thm:main stability thm} guarantees that

\begin{align}
    \|\mathbf{mle}(C)-\mathbf{mle}(D)\|_1&\le 2\log(n^C\lor n^D)W_\infty(C,D) - (L^C\land L^D)2r_\infty(C,D)\log{2r_\infty(C,D)},\label{equ:midlife entropy Winfty bound}\\
    \|\mathbf{mle}(C)-\mathbf{mle}(D)\|_1&\le  \frac{2}{e}W_1(C,D) - (L_\infty^C\land L_\infty^D)2(n^C\lor n^D)r_1(C,D)\log{2r_1(C,D)}. \label{equ:midlife entropy W1 bound}
\end{align}

\section{Proof of \Cref{lem:lemma for pl} in \Cref{subsec:bounds on PL}}
\label{app:landscapes}
For convenience, we will restate the lemma here:
\begin{lemma*}
Let $f(t) =  | \psi_1(t) \chi_{[b_1,d_1)}(t) -\psi_2(t) \chi_{[b_2,d_2)}(t) |$, where $\psi_i(t) = \min\{ t-b_i, d_i - t \}$ for $i = 1,~2$.  Then
\begin{equation*}
    \|f\|_{\infty} \leq |d_2-d_1| \lor |b_2 - b_1|.
\end{equation*}
\end{lemma*}
\begin{proof}
We need similar estimates to that in \Cref{lem:estimate}.  Consider \[f(t) = | \psi_1(t) \chi_{[b_1,d_1)}(t) -\psi_2(t) \chi_{[b_2,d_2)}(t) |,\] where $\psi_i(t) = \min\{ t-b_i, d_i - t \}$ for $i = 1,~2$.  Recall that $\psi_i(t)$ can also be expressed as
\[\psi_i(t) = 
\begin{cases}
0 & \hbox{if } t\notin (b_i,d_i)\\
t-b_i & \hbox{if } t\in [b_i,\frac{b_i+d_i}{2})\\
d_i-t & \hbox{if } t\in[\frac{b_i+d_i}{2}, d_i)
\end{cases}.\]

\noindent Finally, define $m_i = \frac{b_i+d_i}{2}$.

\begin{tikzpicture}
\begin{axis}[
    title = {Case 1},
    axis y line=none,
    y=0.5cm/1.5,
    xtick={5, 9, 11, 13}, 
    xticklabels={$b_1$, $d_1$, $b_2$, $d_2$},
    restrict y to domain=0:1,
    axis lines=left,
    enlarge x limits=upper,
    scatter/classes={
        o={mark=*,fill=white}
    },
    scatter,
    scatter src=explicit symbolic,
    every axis plot post/.style={mark=*,thick},
    legend style={
        draw=none,
        at={(1,1)},
        anchor=south east
    },
    legend image post style={mark=none}
]
\addplot table [y expr=0,meta index=1, header=false] {
5 c
9 o
};
\addplot table [y expr=1,meta index=1, header=false] {
11 c
13 o
};
\end{axis}
\end{tikzpicture}

\noindent Case 1: $b_1 \leq d_1 \leq b_2 \leq d_2$ and $[b_1, d_1) \cap [b_2, d_2) = \emptyset$.  Then
\begin{equation}
    \begin{cases}
    \max_{t\in [b_1,d_1)} | \psi_1(t) | = (d_1 - b_1)/2 \leq (b_2 - b_1)/2 \\
    \max_{t\in [b_2,d_2)} | \psi_2(t) | = (d_2 - b_2)/2 \leq (d_2 - d_1)/2
    \end{cases}.
\end{equation}
Thus, $f(t) \leq |b_2-b_1| \lor |d_2 - d_1|$.

\begin{tikzpicture}
\begin{axis}[
    title = {Case 2},
    axis y line=none,
    y=0.5cm/1.5,
    xtick={5, 7, 9, 13}, 
    xticklabels={$b_1$, $b_2$, $d_1$, $d_2$},
    restrict y to domain=0:1,
    axis lines=left,
    enlarge x limits=upper,
    scatter/classes={
        o={mark=*,fill=white}
    },
    scatter,
    scatter src=explicit symbolic,
    every axis plot post/.style={mark=*,thick},
    legend style={
        draw=none,
        at={(1,1)},
        anchor=south east
    },
    legend image post style={mark=none}
]
\addplot table [y expr=0,meta index=1, header=false] {
5 c
9 o
};
\addplot table [y expr=1,meta index=1, header=false] {
7 c
13 o
};
\end{axis}
\end{tikzpicture}

\noindent Case 2: $b_1 \leq b_2 \leq d_1 \leq d_2$ and $[b_1, d_1) \cap [b_2, d_2) = [b_2, d_1)$.  Then we have three maximums to consider.
\begin{equation}
    \begin{cases}
    \max_{t\in [b_1,b_2]} | \psi_1(t) |   \\
    \max_{t\in [b_2,d_1)} | \psi_1(t) - \psi_2(t) |  \\
    \max_{t\in (d_1,d_2)} | \psi_2(t) |
    \end{cases}.
\end{equation}
For the first, there are two subcases: $b_1 \leq m_1 \leq b_2$ and $b_1 \leq b_2 \leq m_1$. In either case we see 
\begin{equation}
    \begin{cases}
    \max_{t\in [b_1,m_1]} t - b_1 = m_1 - b_1 = \frac{(d_1-b_1)}{2}   \\
    \max_{t\in [m_1,b_2)} d_1 - t = d_1 - m_1 = \frac{(d_1-b_1)}{2}  \\
    \max_{t\in [b_1,b_2)} t - b_1 = b_2 - b_1  
    \end{cases}
\end{equation}
Observe that for $b_1 \leq m_1 \leq b_2$, we have $\frac{(d_1-b_1)}{2} - (b_2 - b_1) = m_1 - b_2 \leq 0 $.  In each of the cases above, $\max_{t\in [b_1,b_2]} | \psi_1(t) |\le b_2-b_1$.    Thus, $\max_{t\in [b_1,b_2]} | \psi_1(t) | \leq |b_2 - b_1|$ .
 Similarly, one may verify that when $t\in (d_1,d_2)$,
$\max_{t\in (d_1,d_2)} | \psi_1(t) | \leq |d_2 - d_1|$.

When $t\in [b_2,d_1) = [b_1, d_1) \cap [b_2, d_2)$, there are 5 sub cases to consider.
\begin{itemize}
    \item[i)] $m_1 \leq b_2 \leq d_1 \leq m_2$;
    \item[ii)] $b_2 \leq m_1 \leq d_1 \leq m_2$;
    \item[iii)] $m_1 \leq b_2 \leq m_2 \leq d_1$;
    \item[iv)] $b_2 \leq m_1 \leq m_2 \leq d_1$;
    \item[v)] $b_2 \leq m_2 \leq m_1 \leq d_1$.
\end{itemize}
One may verify that each case is bounded by $|d_2 - d_1| \lor |b_2 - b_1|$.

\begin{tikzpicture}
\begin{axis}[
    title = {Case 3},
    axis y line=none,
    y=0.5cm/1.5,
    xtick={5, 7, 9, 13}, 
    xticklabels={$b_1$, $b_2$, $d_2$, $d_1$},
    restrict y to domain=0:1,
    axis lines=left,
    enlarge x limits=upper,
    scatter/classes={
        o={mark=*,fill=white}
    },
    scatter,
    scatter src=explicit symbolic,
    every axis plot post/.style={mark=*,thick},
    legend style={
        draw=none,
        at={(1,1)},
        anchor=south east
    },
    legend image post style={mark=none}
]
\addplot table [y expr=0,meta index=1, header=false] {
5 c
13 o
};
\addplot table [y expr=1,meta index=1, header=false] {
7 c
9 o
};
\end{axis}
\end{tikzpicture}

\noindent Case 3: $b_1 \leq b_2 \leq d_2 \leq d_1$. We have three maximums to consider.

\begin{equation}
    \begin{cases}
    \max_{t\in [b_1,b_2]} | \psi_1(t) |   \\
    \max_{t\in [b_2,d_2)} | \psi_1(t) - \psi_2(t) |.  \\
    \max_{t\in (d_2,d_1)} | \psi_1(t) |
    \end{cases}
\end{equation}
The first and third maximums follow the same arguments as in case 2. For the middle maximum, we see that we have four subcases

\begin{itemize}
    \item[i)] $m_1\in (b_1,b_2]$;
    \item[ii)] $m_1\in(b_2,m_2]$;
    \item[iii)] $m_1\in(m_2,d_2)$;
    \item[iv)] $m_1\in(d_2,d_2)$.
\end{itemize}

For i), note that $d_1-b_2 < d_1-m_1 = m_1-b_1$ thus, $d_1-b_2 < b_2-b_1$. We complete this case by noting that for all $t\in[b_2,d_2)$, $\psi_1(t) = d_1-t \ge \psi_2(t)\ge 0$. Moreover $\psi_1$ is decreasing over the interval $[b_2,d_2)$. Therefore, the difference $|\psi_1(t) - \psi_2(t)|$ is maximized at $t=b_2$ and the difference is precisely $d_1-b_2$.

One may verify the remaining cases satisfy the conclusion.
\end{proof}
\end{document}